\documentclass{article}

\pdfoutput=1
\usepackage{arxiv}
\usepackage{verbatim}

\usepackage[utf8]{inputenc} 
\usepackage[T1]{fontenc}    
\usepackage{hyperref}       
\usepackage{url}            
\usepackage{booktabs}       
\usepackage{amsfonts}       
\usepackage{nicefrac}       
\usepackage{microtype}      
\usepackage{lipsum}

\usepackage{cite}
\usepackage[pdftex]{graphicx}
\usepackage{subcaption}
\usepackage{epstopdf}
\usepackage{tikz-cd}
\usepackage{amsmath,amssymb,amsfonts}
\usepackage{bm} 
\newcommand{\mycircle}[1]{\raisebox{.5pt}{\textcircled{\raisebox{-.9pt}{#1}}}}

\title{A Network-Guided Reaction-Diffusion Model of AT[N] Biomarkers in Alzheimer’s Disease}

\author{
  Jingwen Zhang\\
  Department of Computer Science\\
  Wake Forest University\\
  Winston-Salem, NC 27109 \\
  \texttt{zhanj318@wfu.edu} \\
  \And
 Defu Yang\\
 Department of Psychiatry \\
 UNC at Chapel Hill\\
 Chapel Hill, NC 27599\\
 \texttt{dfyang@hdu.edu.cn}
\And
 Wei He\\
 Genetics, Bioinformatics, and\\
 Computational Biology. Virginia Tech\\
 Blacksburg, VA 24061 \\
 \texttt{hwei9@vt.edu}
\And
 Guorong Wu\\
 Department of Psychiatry, Computer Science \\
 University of North Carolina at Chapel Hill\\
 Chapel Hill, NC 27599 \\
 \texttt{grwu@med.unc.edu}
 \And
  Minghan Chen\\
  Department of Computer Science\\
  Wake Forest University\\
  Winston-Salem, NC 27109 \\
  \texttt{chenm@wfu.edu} \\
}

\begin{document}

\maketitle

\begin{abstract}
Currently, many studies of Alzheimer's disease (AD) are investigating the neurobiological factors behind the acquisition of beta-amyloid (A), pathologic tau (T), and neurodegeneration ([N]) biomarkers from neuroimages. However, a system-level mechanism of how these neuropathological burdens promote neurodegeneration and why AD exhibits characteristic progression is largely elusive.
In this study, we combined the power of systems biology and network neuroscience to understand the dynamic interaction and diffusion process of AT[N] biomarkers from an unprecedented amount of longitudinal Amyloid PET scan, MRI imaging, and DTI data. Specifically, we developed a network-guided biochemical model to jointly (1) model the interaction of AT[N] biomarkers at each brain region and (2) characterize their propagation pattern across the fiber pathways in the structural brain network, where the brain resilience is also considered as a moderator of cognitive decline.
Our biochemical model offers a greater mathematical insight to understand the physiopathological mechanism of AD progression by studying the system dynamics and stability.
Thus, an in-depth system-level analysis allows us to gain a new understanding of how AT[N] biomarkers spread throughout the brain, capture the early sign of cognitive decline, and predict the AD progression from the preclinical stage.

\end{abstract}


\section{Introduction}

With the rapid development of neurobiology and neuroimaging technologies, mounting evidence shows that Alzheimer’s disease (AD) is caused by the build-up of two abnormal proteins, beta-amyloid and pathologic tau. 
Over time, these AD-related neuropathological burdens begin to spread throughout the brain, which results in the characteristic progression of symptoms in AD. 

According to the National Institute of Aging and Alzheimer’s Association (NIA-AA) Research Framework~\cite{jack2018nia},
the biological hallmarks of AD can be grouped and classified as: beta-amyloid (A), pathologic tau (T), and neurodegeneration ([N]). When amyloid precursor protein (APP), the transmembrane protein, is cut by beta and gamma secretases, a peptide called beta-amyloid (A$\beta$) will be released~\cite{lloret2015molecular}. The aggregates of A$\beta$ form insoluble plaques among neurons which can 
depolarize synapse membrane, trigger inflammatory responses, and induce synaptic loss~\cite{spires2014intersection, chen2017amyloid, von2007pro, gamba2011interaction}. Recognized as the early sign of AD~\cite{leal2018subthreshold},  A$\beta$ is widely studied in both theoretical and clinical research. Another hallmark tau biomarker, a microtubule binding protein, is formed by alternative splicing controlled by gene MAPT~\cite{caillet2015regulation}. When hyperphosphorylated, tau will disassociate from axons and form neurofibrillary tangles (NFT), causing synaptic dysfunction and neuron damage~\cite{mandelkow2012biochemistry, rademakers2004role, quintanilla2014phosphorylated}. Neurodegeneration, the downstream biomarker of A and T, links directly with brain atrophy and is considered as the final signal of a series of cognitive disorders such as AD when deteriorating to a certain level. 
 
In the past few decades, several neuropathologies 
have been proposed to explain the whole-brain progression of AD. One of the main pathologies is the amyloid cascade hypothesis~\cite{hardy1992alzheimer, chen2017amyloid, spires2014intersection}, in which the fibrillar form of A$\beta$ aggregates into senile plaques and activates the hyperphosphorylation process of tau protein~\cite{lloret2015molecular, mandelkow2012biochemistry, manczak2013abnormal}. The hyperphosphorylated tau causes neurodegeneration and eventually leads to AD.
Recently, growing evidence in culture models and animal models favors the oligomeric amyloid hypothesis and suggests oligomeric amyloid as a central pathogenic role in AD progression~\cite{lauren2009cellular, balducci2010synthetic}. The prion-like oligomeric form of A$\beta$ interrupts regular synaptic function, causes synapse shrinkage~\cite{cleary2005natural, selkoe2008soluble}, and induces long-term depression (LTD)~\cite{spires2014intersection}. The other tauopathy hypothesis emphasizes the functionality of biomarker T and suggests that the detachment of hyperphosphorylated tau from microtubule and deposition of neurofibrillary tangles is the leading cause of neuron dysfunction~\cite{chen2017amyloid, mandelkow2012biochemistry, jack2018nia, lloret2015molecular}.

Based on the existing hypothesis, biochemical models with different focuses and scales have been built to study the biomarker dynamics and disease progression. One concentration is to model the kinetics of fibrillation process of beta-amyloid on molecular level~\cite{cohen2013proliferation, lloret2017impact, qosa2014differences, proctor2012aggregation,pallitto2001mathematical}. 
Pallito and Murphy described the beta-amyloid fibril growth in a progressive and comprehensive mechanism as monomer addition, fibril association, filament-fibril aggregation~\cite{pallitto2001mathematical}.
Another focus is to construct a systematic model by investigating the interactions between different biomarkers and probing into a more detailed mechanism behind this multifactorial neurodegenerative disease~\cite{dunster2014resolution, puri2010mathematical, jack2010hypothetical}. For example, Hao and Friedman proposed a schematic network of AD that models the interactions between beta-amyloid, neurofibrillary tangles, astrocytes, microglia, and macrophages~\cite{hao2016mathematical}. 
A trending direction nowadays is to extend the model to the real whole-brain scale, evaluating the prion-like evolution and diffusion of different misfolded peptides or proteins~\cite{ helal2014alzheimer, weickenmeier2018multiphysics, iturria2014epidemic}. Among them, Raj et al. proposed a network diffusion model for regional atrophy using the brain's connectivity network derived from imaging data~\cite{raj2012network,kim2019constructing}; the model was further developed and tested by empirical magnetic resonance imaging (MRI) and fluorodeoxyglucose (FDG) positron emission tomography (PET) to predict end-of-study atrophy pattern ~\cite{raj2015network}. 
 

As neuroimaging data become increasingly available, most approaches proposed today utilize the statistical inference to understand the neurobiological factors behind AD progression.
However, these approaches only address ``mechanisms" via proxies that are often distinct from the essential pathophysiological mechanisms.
On the other hand, most AD-related systems biology approaches are limited to studying a single pathology pathway or a small part of the brain and lack the whole-brain insight gained from longitudinal neuroimaging data. In this work, we propose a novel network-guided biochemical model to understand the reaction-diffusion process of AT[N] biomarkers across brain networks. First, we use a set of partial differential equations (PDEs) to model the interactions between AT[N] biomarkers in each brain region. The neuropathological burdens are allowed to diffuse along the fiber pathways in the structural brain network in a prion-like manner. Second, we investigate the system behavior of our network-guided biochemical model by applying a classic stability analysis, 
where system-level patterns emerge as a putative AD biomarker.



\section{Model}
In this section, we adapt a well-studied bistable model to characterize the dynamic neurodegeneration process in the whole brain. We hypothesize that our network-guided biochemical model can interpret the complicated (sometimes even controversial) relationship between spatiotemporal progression of neuropathological burdens and the resulting cognitive decline in the aging population. 

\subsection{A network-guided bistable model}

\begin{figure*}[!htb]
\begin{subfigure}{0.49\textwidth}
\centering
\includegraphics[width=0.78\linewidth]{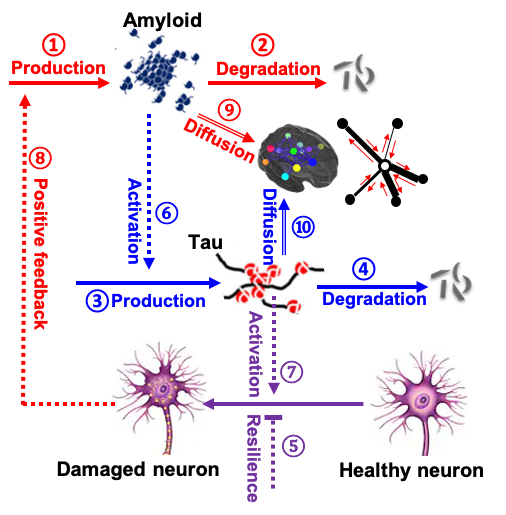}
\caption{Schematic representation of AD pathological network}
\label{fig:diagram}
\end{subfigure}
\begin{subfigure}{0.49\textwidth}
\centering
\includegraphics[width=0.9\linewidth]{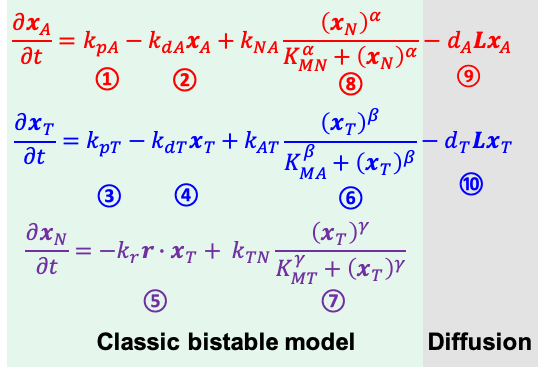}
\caption{Mathematical representation of AD pathological network}
\label{fig:equation}
\end{subfigure}
\caption{\textbf{Overview of our network-guided biochemical model which characterizes the reaction-diffusion process of AT[N] biomarkers across brain network by integrating the bistable model (green) and network diffusion (grey).}
{ Each numbered pathway in \textbf{(a)} corresponds to the numbered term in \textbf{(b)}. 
The constant production, degradation, and diffusion terms for $\beta$ amyloid and pathological tau were proposed, and three hill equations were used to model the interactions between AT[N] biomarkers based on the cascade: A stimulates the production of pathological T, which, in turns, activate neurodegeneration. Neuronal resilience was considered and a positive feedback between A and [N]. 
First, in each region of the brain, we consider the evolution of A and T biomarkers following natural production (\mycircle{1}, \mycircle{3}) and degradation (\mycircle{2}, \mycircle{4}) pathways. Second, we include the activation and positive feedback pathways among A, T, and [N] biomarkers. Specifically, the activation pathways (\mycircle{6}, \mycircle{7}) follow the $\rm A \rightarrow T \rightarrow [N]$ mainstream that amyloid induces the spread of pathological tau and tau is immediately proximate to neurodegeneration, i.e., amyloid cascade hypothesis. The positive feedback pathway (\mycircle{8}) describes the effect that damaged neurons may stimulate the production of amyloid via reactive astrocytes. Third, we specifically take the subject’s network resilience (in terms of the moderated ratio of neuron loss) by modeling an inhibition pathway (\mycircle{5}). Last, since A and T biomarkers diffuse along the topological path in a prion-like manner, we employ graph Laplacian matrix $L$ as the diffusion operator on the brain network (\mycircle{9}, \mycircle{\small 10}). 
}
}
\label{fig:model}
\end{figure*}

The overview of our schematic network is presented in Fig.~\ref{fig:diagram}. Since there is no conclusive evidence favoring one specific form of amyloid over the others, we assume amyloid biomarker represents the A$\beta$ protein of all formats, including monomers, oligomers, polymers, aggregated fibrils, and plaques. Similarly, tau biomarker consists of hyperphosphorylated tau monomers, oligomers, polymers, aggregated fibrils, and neurofibrillary tangles. The final entity we modeled is the neurodegeneration biomarker, an indicator of the neural damage caused by abnormal activities of both A and T. We assume soluble amyloid and tau can diffuse along the topological path in the brain network~\cite{chen2017amyloid,raj2015network,spires2014intersection}. 

Based on the proposed mechanistic pathways described in Fig.~\ref{fig:model}, our model consists of three sets of PDEs that characterize the spatiotemporal dynamics for AT[N] biomarkers, respectively. In the green box of Fig.~\ref{fig:equation}, the first set of equations describes the concentration change of amyloid across the brain (denoted as ${\bm x}_A$) as a result of natural production, degradation, and positive feedback from damaged neurons. 
$k_{pA}$ is the rate constant of amyloid production through the proteolysis process of APP. The degradation of amyloid includes active transport to blood that acrosses blood-brain barrier, proteolytic degradation controlled by proteases, and microglia mediated clearance. $k_{dA}{\bm x}_A$, proportional to amyloid concentration, is used to represent the above degradation pathways in general. When the neuron is damaged or dead, the apoptosis process will release intracellular amyloid to the brain~\cite{hao2016mathematical}, and stimulate the expression of APP and A$\beta$~\cite{siman1989expression, otsuka1991rapid}, both leading to a positive feedback loop between A and [N]. 
Since the underlying physiological mechanism of the positive feedback is still unclear and likely nonlinear, the classic Hill function~\cite{weiss1997hill, goutelle2008hill} is applied to approximate the molecular processes as nonlinear reactions, where $k_{NA}$ is the positive feedback constant, $K_{MN}$ is the dissociation constant, and $\alpha$ is the Hill coefficient.

Similar to the form of amyloid, the second set of equations describes the concentration change of tau over time across the brain (denoted as ${\bm x}_T$) with its own set of rate constants $k_{pT}$, $k_{dT}$, $k_{AT}$, $K_{MA}$, $\beta$. 
The last set of equations models the change in neural damage across the brain (denoted as ${\bm x}_N$) resulting from tau activation and brain resilience. Brain resilience is hypothesized to preserve cognition despite underlying AD pathology~\cite{stern2017approach}
and is steered by the term $k_r {\bm r}$, where ${\bm r}$ is the average resistance force of degeneration at each brain region calculated from ~\cite{Xie2020Characterizing} and $k_r$ is the corresponding rate constant. We use the Hill function for the activation pathway ($\rm T \rightarrow [N] $) as the physiological process is likely nonlinear with rate constants $k_{TN}$, $K_{MT}$, $\gamma$.

In the grey box of Fig.~\ref{fig:equation}, the diffusions of amyloid and tau are guided by the Laplacian matrix $L$ of average brain network. Suppose we parcellate the whole brain into $M$ regions and render a $M \times M$ adjacency matrix $W$ which encodes region-to-region connectivity strength. 
We estimate the average adjacency matrix $\overline W$ across subjects.
Then $L$ can be obtained by $L=D-\overline W$, where $D$ is a diagonal matrix with each diagonal element equals to the total connectivity degree of the underlying node.
Thus, $d_AL{\bm x}_A$ and $d_TL{\bm x}_T$ characterize the dynamic balance of influx and outflux of neuropathological burdens at each node, where $d_A$ and $d_T$ are the diffusion rates of amyloid and tau. Note that ${\bm x}_A$, ${\bm x}_T$, and ${\bm x}_N$ are column vectors of size $M$ (e.g., ${\bm x}_A=\{ x_{A1}, x_{A2}, ..., x_{AM}\}$), representing the degree of AT[N] biomarkers at each region. In this work, we use hyperparameter $\Theta$ to denote the total 16 model parameters, which will be fitted with the longitudinal observation of AT[N] biomarkers.



\subsection{System behavior analysis}
The green part of the proposed biochemical model is also a typical bistable model. Integrating the bistability feature with network diffusion, model results will be driven into two system states. 
We first investigate model equilibria, at which the system does not change over time. Take the PDE system $\frac{\partial x}{\partial t} =f(x)$ as an example, the equilibrium point is the solution of $f(x^*)=0$, which is also known as the steady-state of the system. 
We further perform a stability analysis to find stable equilibrium, which refers to a system that returns to its equilibrium point and remains there after disturbances. Likewise, an unstable equilibrium is a system that moves away from the equilibrium after disturbances. Since our model is a nonlinear dynamic system on a continuous-time domain, we use Lyapunov’s stable theory~\cite{lasalle1961stability} to analyze the local stability of the detected equilibria by solving the characteristic equations of the PDEs. An equilibrium point is stable if and only if the real part of the eigenvalues 
are all negative. 

\begin{figure}[!htb]
\centering
\includegraphics[width=0.4\linewidth]{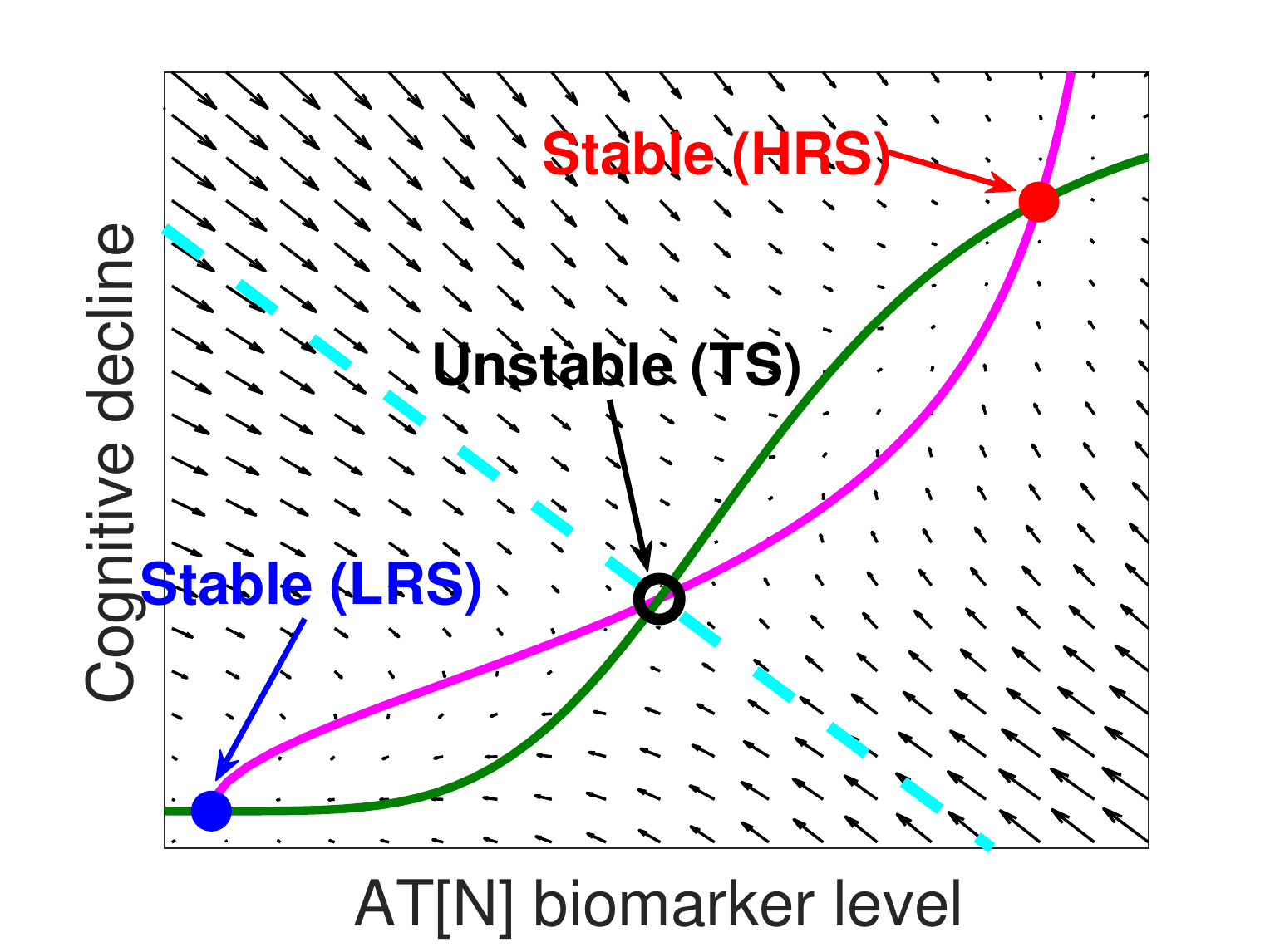}
\caption{\textbf{Demonstration of system stability (phase plane).} 
{ Our bistable model has three equilibrium points, two of which are stable. One stable point with lower AT[N] level corresponds to low risk state (LRS) and the other with higher AT[N] level corresponds to high risk state (HRS). The middle unstable point refers to the transition state (TS) between LRS and HRS. The arrows (vector field) indicate the moving direction towards the system's final state.}}
\label{fig:bistable}
\end{figure}

As a bistable model, we expect to find two stable steady states, differentiated as low and high states. 
Since AD progression is generally non-reversible, it is easy to associate the low steady-state with relatively low level of AT[N] to low risk state (LRS), and the high steady-state with relatively high level of AT[N] to high risk state (HRS). As demonstrated in Fig.~\ref{fig:bistable}, the understanding of the system level reaction-diffusion behavior allows us to predict the trajectory of cognitive decline for aging individuals in two steps: (1) predict the subject-specific AT[N] dynamics by fitting the PDEs using the baseline measurements; (2) identify the subject-specific equilibria $x^*$ and the corresponding stability. As an outcome of our biochemical model, we can predict the possibility of whether the aging individual has a high risk to develop AD or maintain a normal level of cognition.

\section{Results and discussions}

\subsection{Neuroimaging data processing}

\begin{figure}[!htb]
\centering
\begin{subfigure}{0.23\textwidth}
\centering
\includegraphics[width=1\linewidth,height=1.1\linewidth]{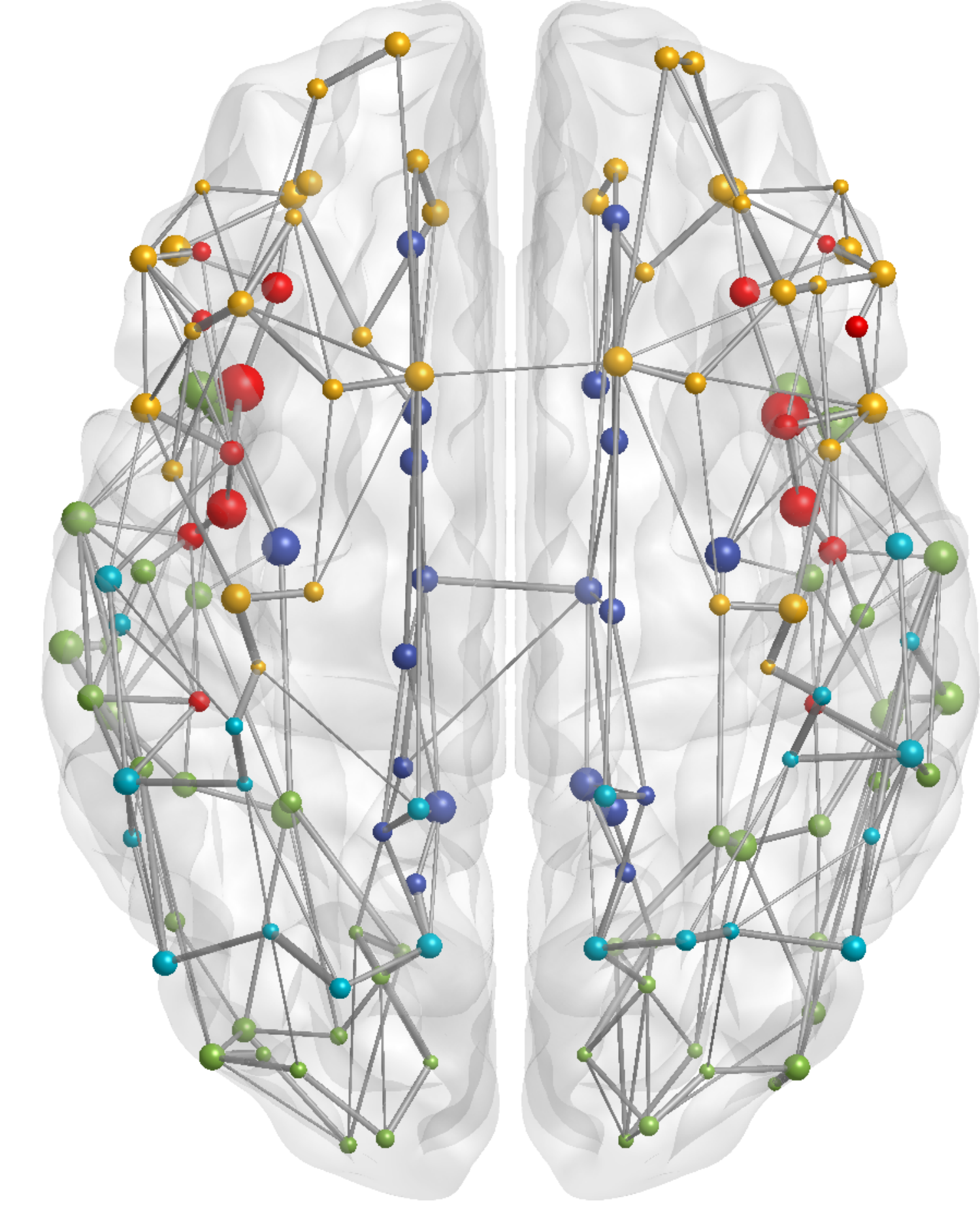}
\caption{Axial View}
\label{fig:edge1}
\end{subfigure}
\begin{subfigure}{0.25\textwidth}
\centering
\includegraphics[width=0.9\linewidth]{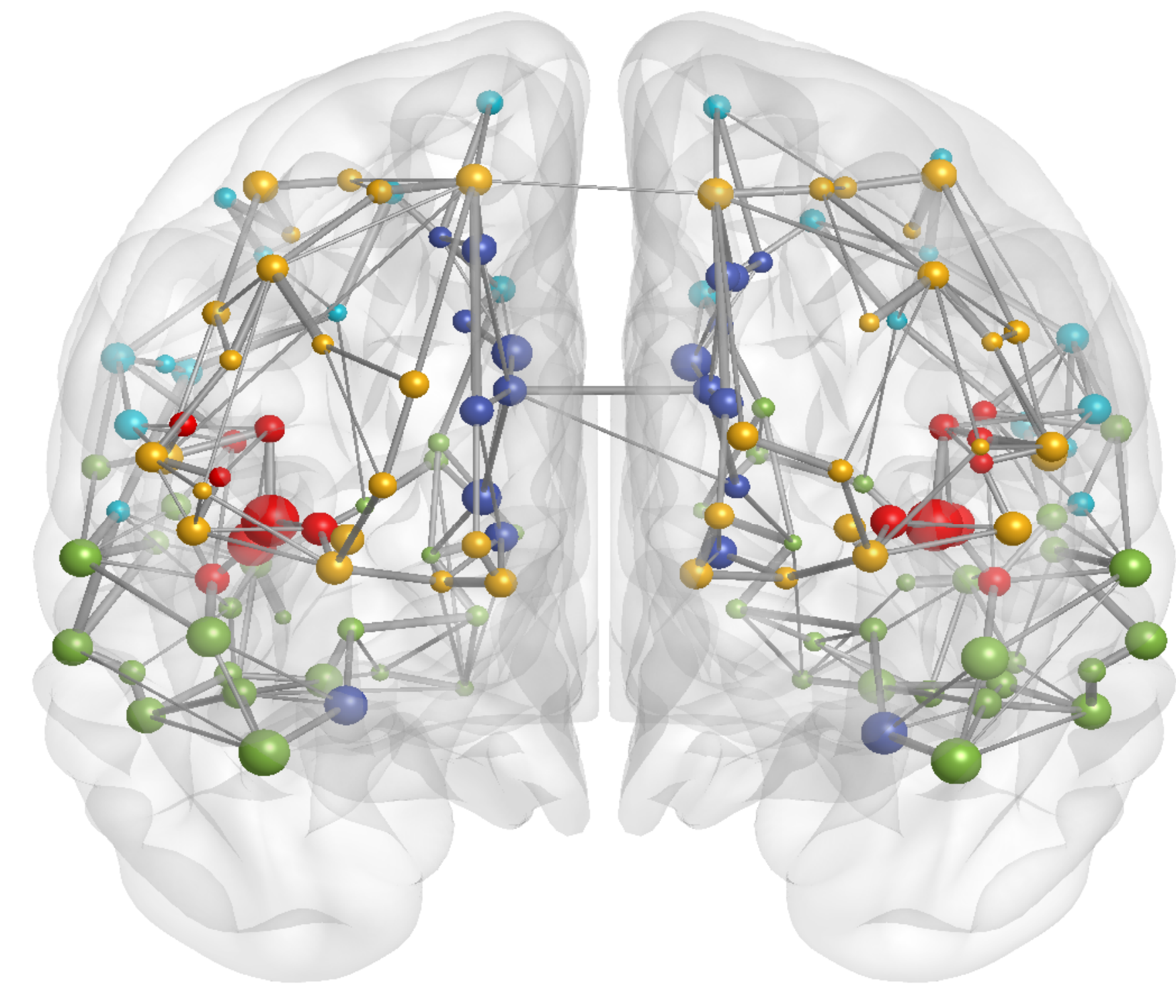}
\caption{Coronal View}
\label{fig:edge2}
\end{subfigure}
\begin{subfigure}{0.3\textwidth}
\centering
\includegraphics[width=0.9\linewidth]{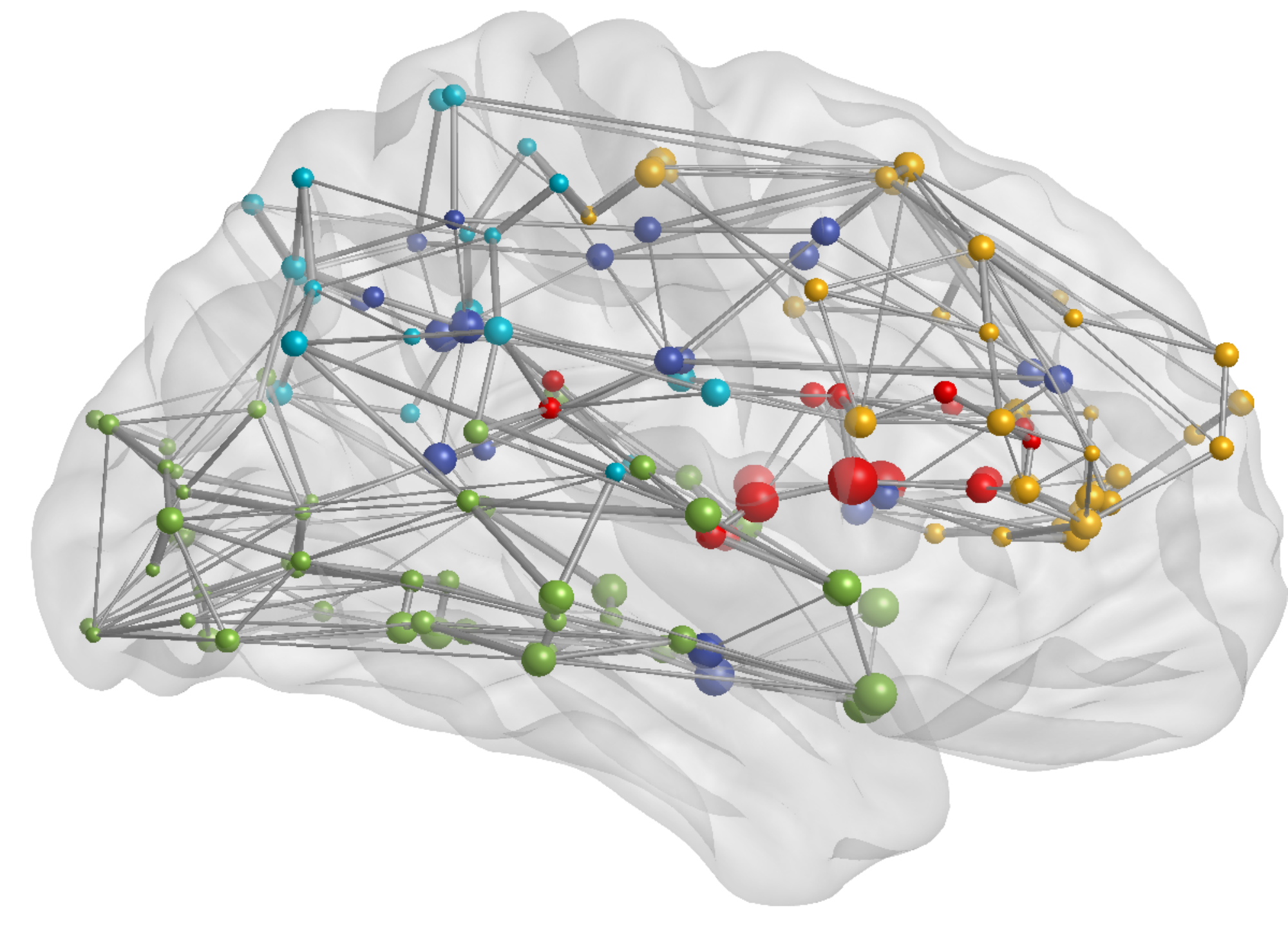}
\caption{Sagittal View}
\label{fig:edge3}
\end{subfigure}
\begin{subfigure}{0.1\textwidth}
\centering
\includegraphics[width=1\linewidth]{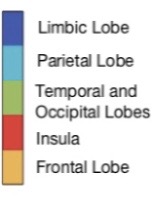}
\end{subfigure}
\caption{\textbf{Brain network and connectivity.} 
{ The whole brain was parcellated into 5 modules and 148 regions.
The color of node denotes the located module. Frontal Lobe (yellow), Insula Lobe (red), Temporal and Occipital Lobes (green), Parietal Lobe (cyan), Limbic Lobe (blue). 
The size of node denotes the cortical thickness at the region derived from a random MRI scan.
Two nodes are connected by an edge (grey) if their connectivity strength is $\ge 0.2$ after normalization 
and the corresponding edge thickness indicates the node-to-node anatomical connectivity strength. Brain images with node representations in this paper were generated via BrainNet Viewer~\cite{xia2013brainnet}.}
}
\label{fig:edge}
\end{figure}

A total of 1897 longitudinal neuroimaging data of 200 cohorts from the Alzheimer's Disease Neuroimaging Initiative (ADNI) were processed and used in this study, including 579 Amyloid PET scans, 812 MRI scans, and 506 diffusion tensor imaging (DTI) scans. Amyloid PET , frequently used in the clinical diagnosis of neurodegenerative diseases, 
allows detection of amyloid plaques in vivo and thus is used as a reference of amyloid level in each brain region~\cite{ossenkoppele2015prevalence, palmqvist2015detailed,wolk2009amyloid}. 
MRI, which measures cortical thickness,
is reversed to represent neuronal loss. DTI is used as an indicator of the connectivity strength of brain network based on tractography on the image~\cite{gong2012convergence, pettigrew2016cortical}.
After partition the cortical surface into 148 regions using Destrieux atlas~\cite{destrieux2010automatic} in free surfer, a set of structured 148 $\times$ 148 network data, regional amyloid and cortical thickness were derived for each subject. Fig.~\ref{fig:edge} presents the average brain network used in our model, where the whole brain was parcellated into 148 regions and the network connectivity was calculated based on DTI scans. Since almost all regions in the brain are connected, the network is a highly dense graph with a large variation of connectivity strength. To reduce the network density, we set up a threshold to filter out weak connections. The edges in Fig.~\ref{fig:edge} show the network connectivity where we cut off connections if the connectivity strength is less than 0.2 after normalization.

\begin{figure}[!htb]
\centering
\begin{subfigure}{0.24\textwidth}
\centering
\includegraphics[width=1\linewidth]{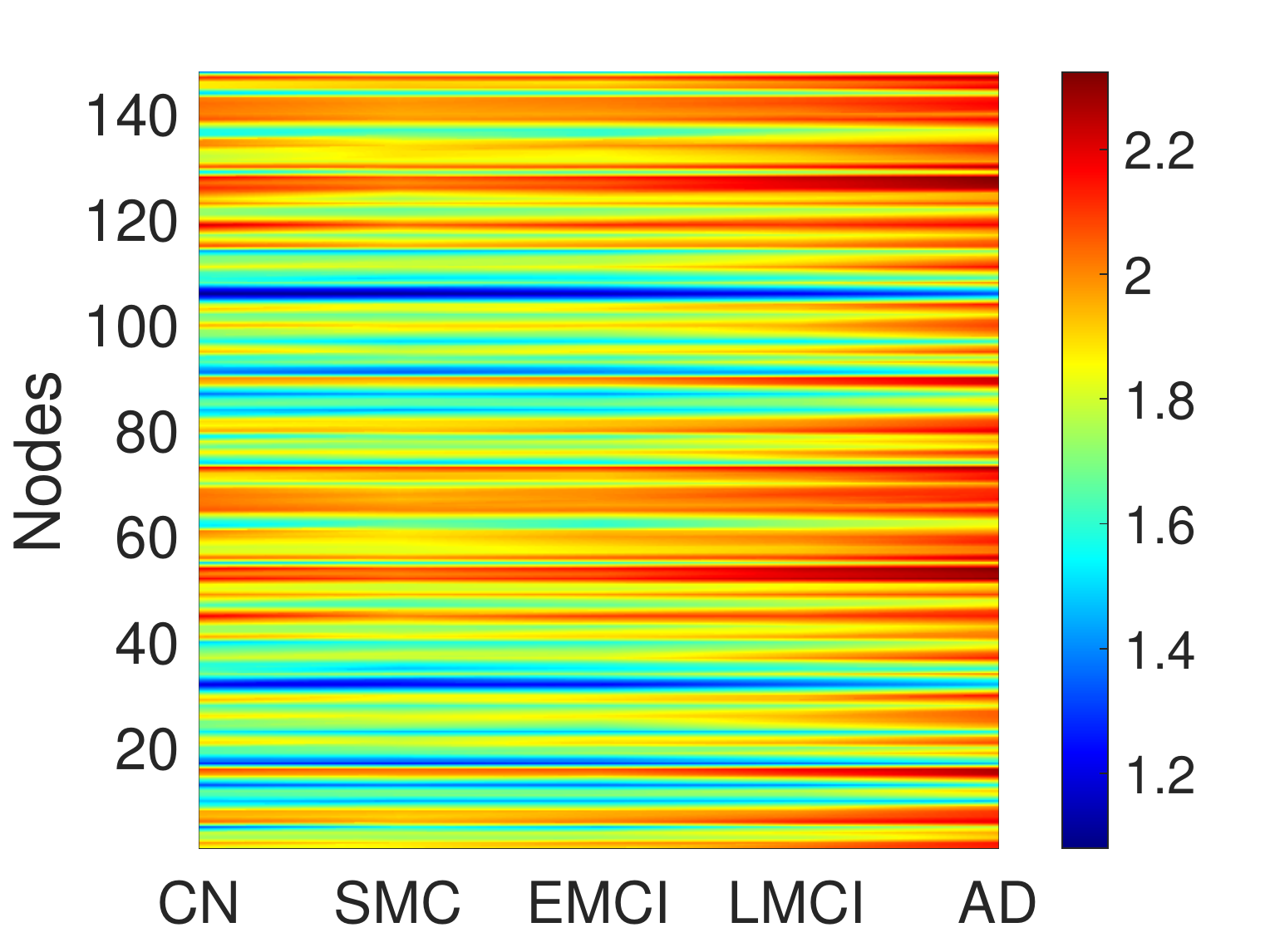}
\caption{Amyloid Level}
\label{fig:PcolorAmy}
\end{subfigure}
\begin{subfigure}{0.24\textwidth}
\centering
\includegraphics[width=1\linewidth]{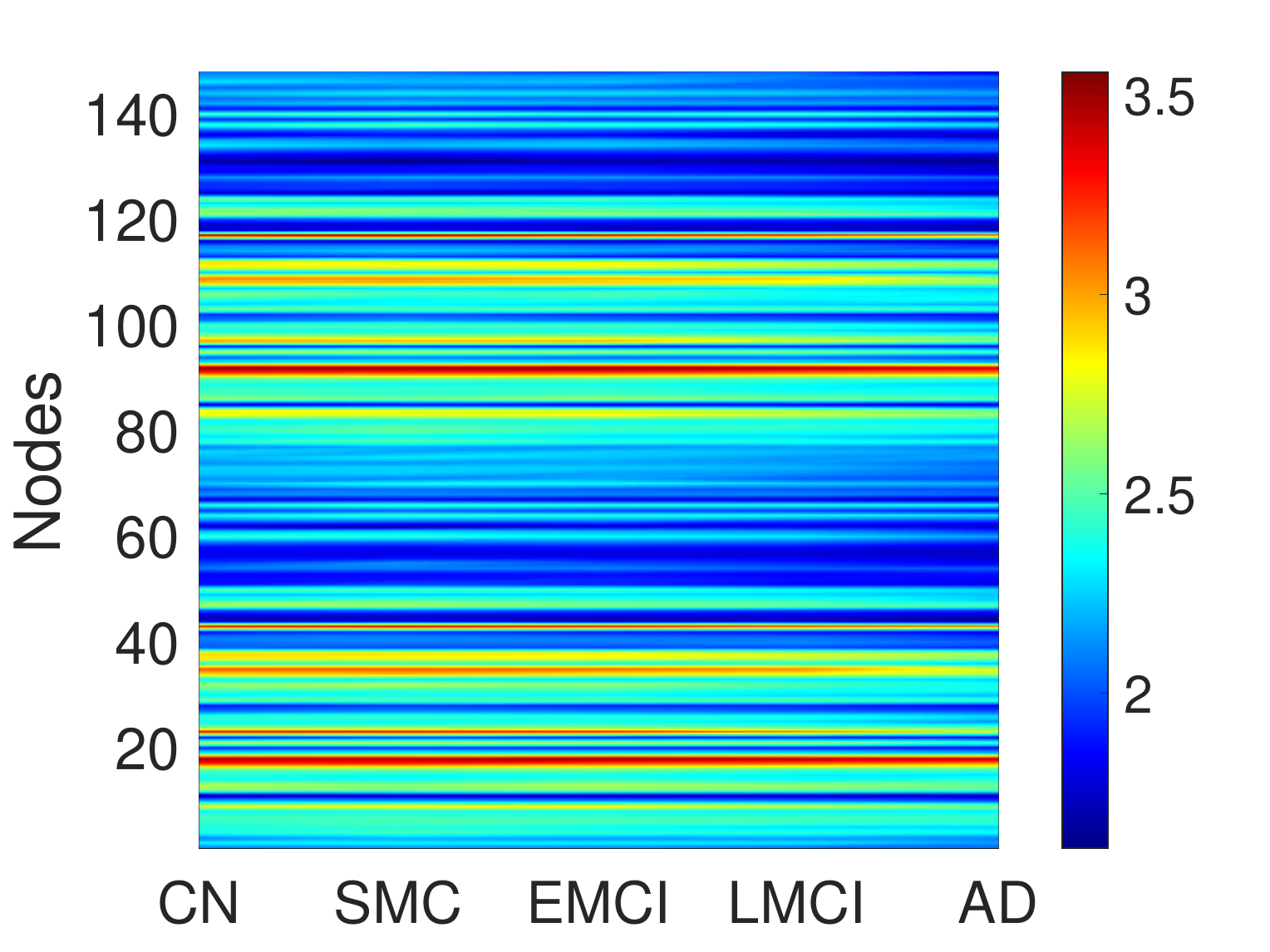}
\caption{Cortical Thickness}
\label{fig:PcolorCor}
\end{subfigure}
\begin{subfigure}{0.24\textwidth}
\centering
\includegraphics[width=1\linewidth]{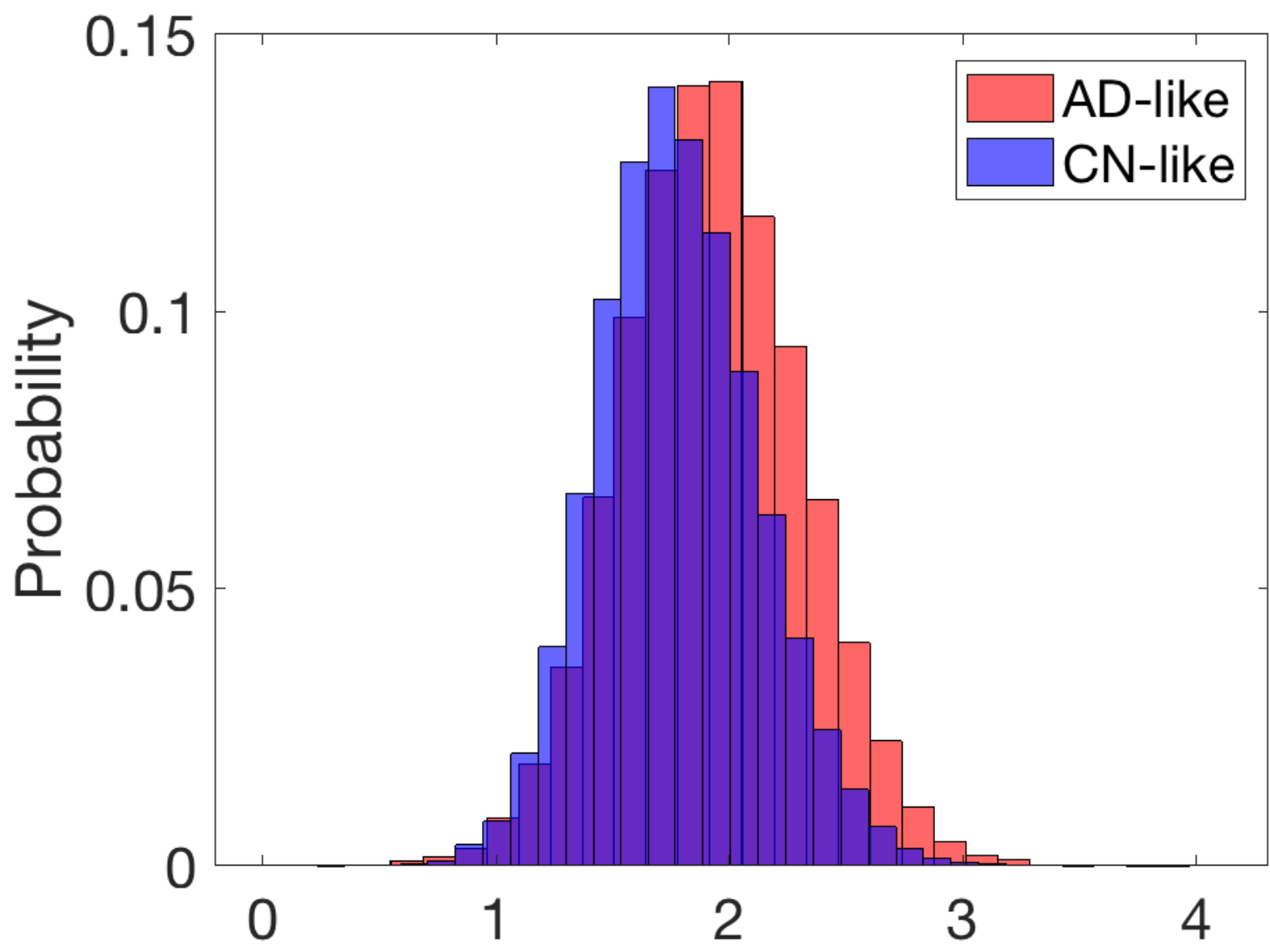}
\caption{Regional Amyloid Level}
\label{fig:amyloidHist}
\end{subfigure}
\begin{subfigure}{0.24\textwidth}
\centering
\includegraphics[width=1\linewidth]{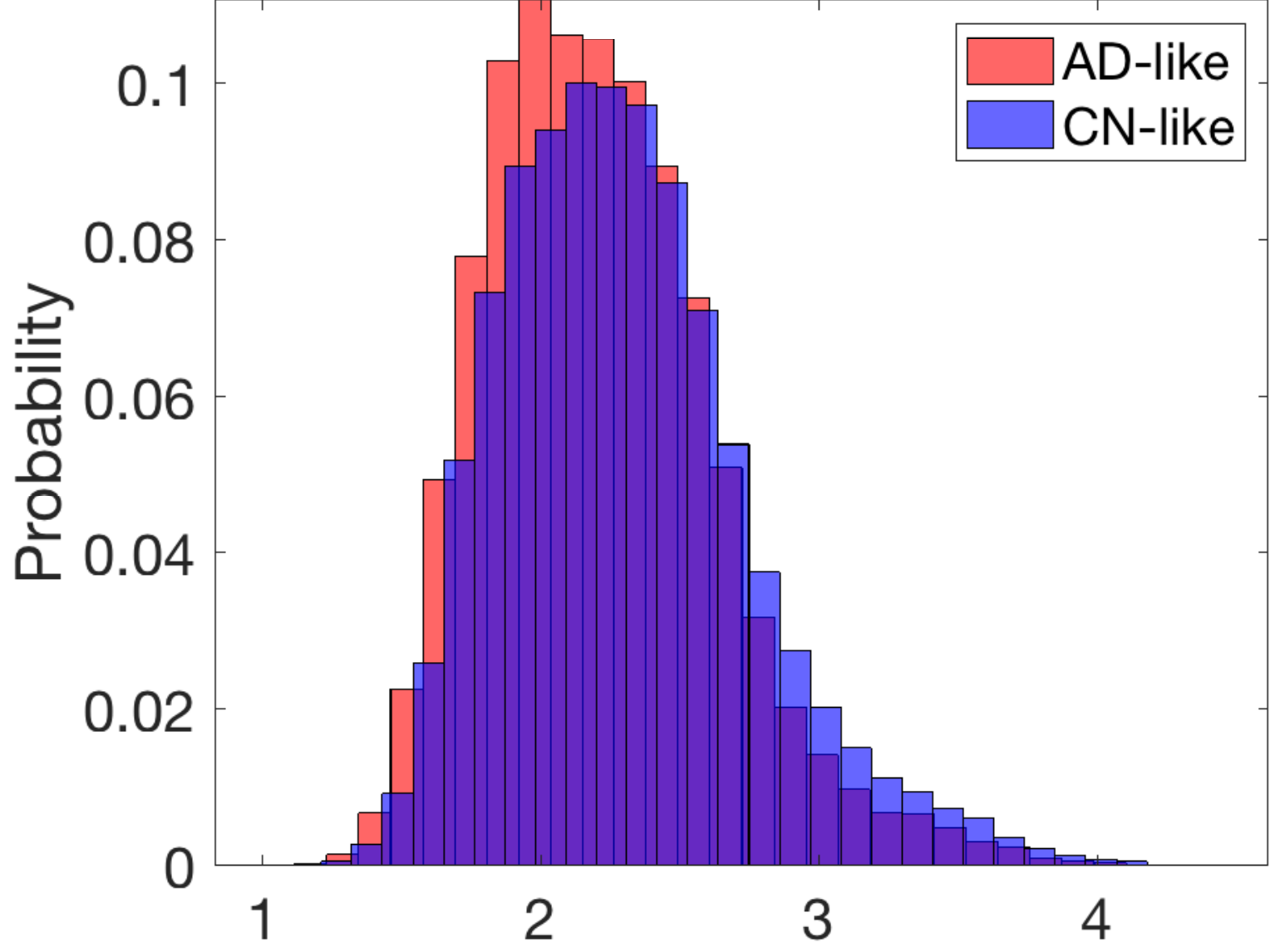}
\caption{Cortical Thickness Level}
\label{fig:corHist}
\end{subfigure}
\caption{\textbf{Statistical Analysis of Amyloid and Cortical Thickness Level.}
\textbf{(a, b)} Heatmaps of amyloid levels and cortical thickness over 148 brain nodes and five diagnostic labels. All subjects' Amyloid PET and MRI data were binned into groups (CN, SMC, EMCI, LMCI, AD) and a nodal mean was calculated.
\textbf{(c, d)} Empirical PDFs of regional amyloid and cortical thickness levels for CN-like and AD-like groups. 
}
\label{fig:data}
\end{figure}

We divided the 200 cohorts into five groups based on the diagnostic labels of their last scan, including 41 subjects in cognitive normal (CN) group, 24 in significant memory concern (SMC), 62 in early-stage mild cognitive impairment (EMCI), 15 in late-stage mild cognitive impairment (LMCI), and 58 in AD. 
Further partition was conducted based on the clear difference in amyloid level as the disease progresses. 
In the heatmaps of A[N] biomarkers over the five groups (Figs.~\ref{fig:PcolorAmy} and~\ref{fig:PcolorCor}), an obvious whole-brain-scale elevation of amyloid was observed between EMCI and LMCI. There is a weak decline in cortical thickness starting from LMCI near the neocortex and hippocampus, such as node 23 (parahippocampal gyrus region), which is consistent with previous observations~\cite{siman1989expression, spires2014intersection, luo2016cross, manczak2013abnormal, chen2017amyloid}. 
We merge CN, SMC, and EMCI groups as `CN-like', and merge LMCI and AD groups as `AD-like'. In the empirical probability density plots of A[N] biomarkers (Fig.~\ref{fig:data}), most of the CN-like and AD-like distributions overlap with each other, with a subtle shift that requires a highly sensitive model to differentiate the two groups.

\subsection{Simulation results and predictions}
\begin{figure*}[!htb]
\centering
\fbox{

\begin{subfigure}{0.46\textwidth}
\centering
\textbf{\large Subject A}

\includegraphics[width=0.4\linewidth,height=0.44\linewidth]{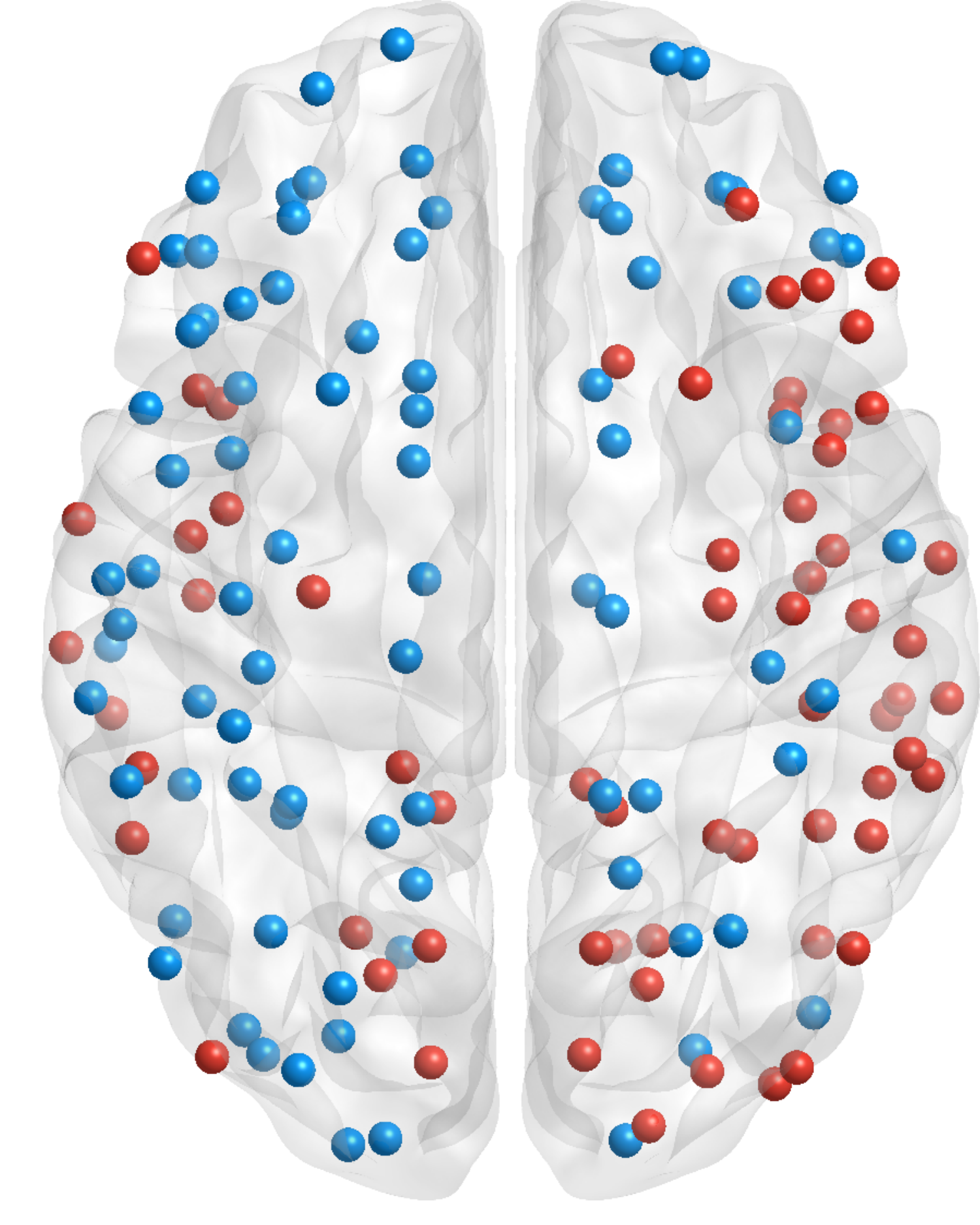}
\hspace{0.01\textwidth}
\includegraphics[width=0.4\linewidth,height=0.444\linewidth]{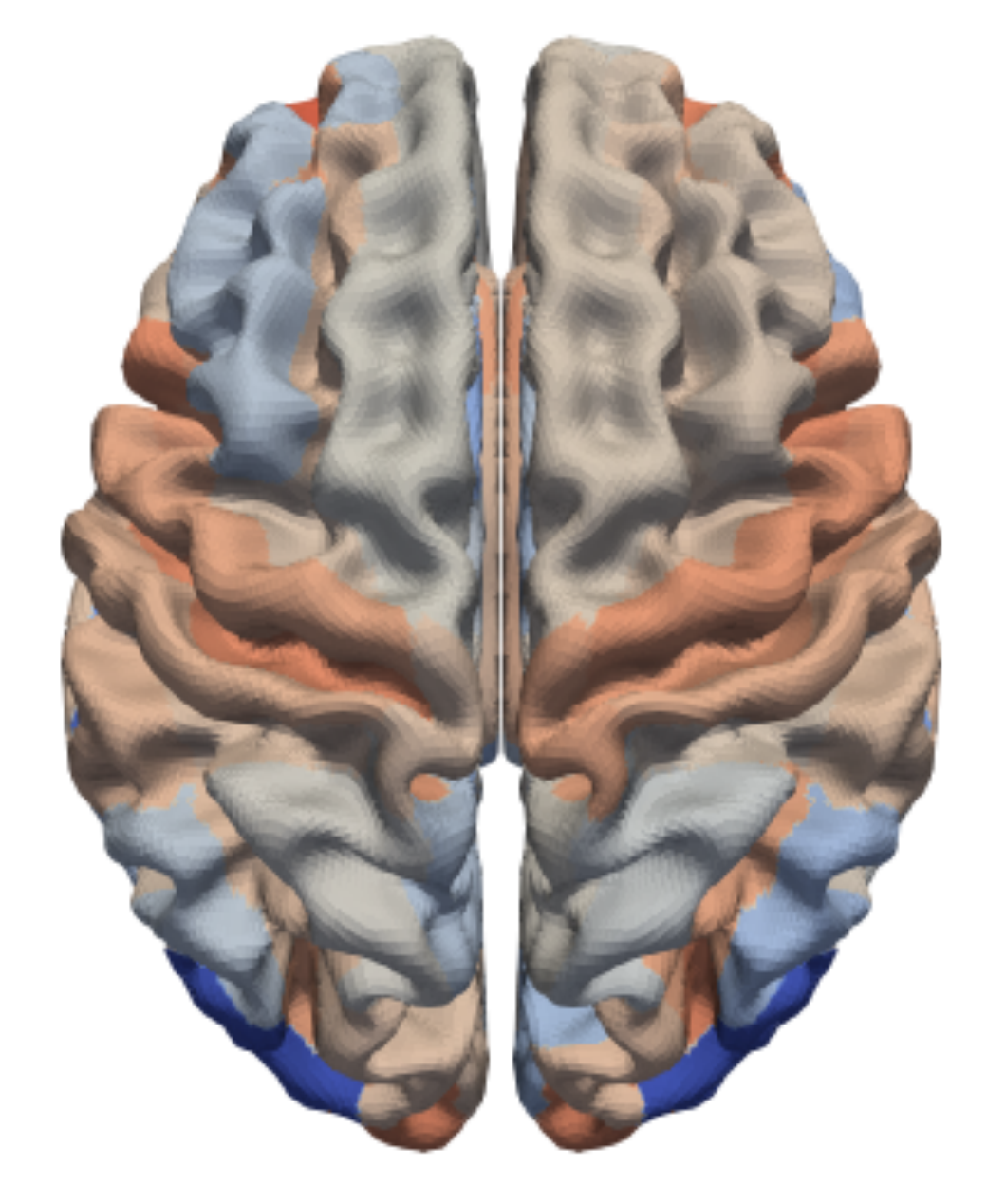}
\caption{Baseline scan of amyloid (left) and thickness (right) as EMCI}

\hspace{0.01\textwidth}

\includegraphics[width=0.4\linewidth,height=0.44\linewidth]{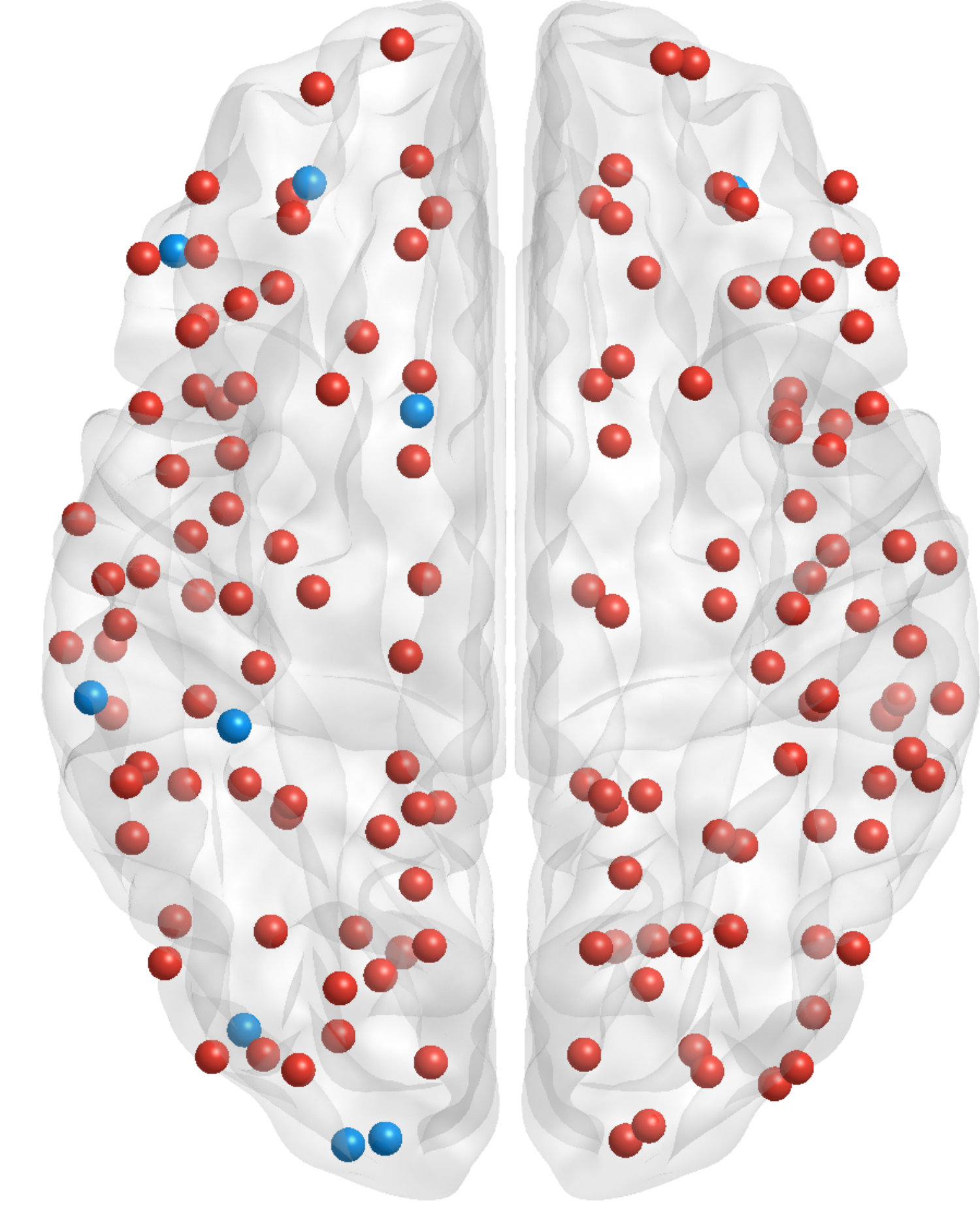}
\hspace{0.01\textwidth}
\includegraphics[width=0.4\linewidth,height=0.44\linewidth]{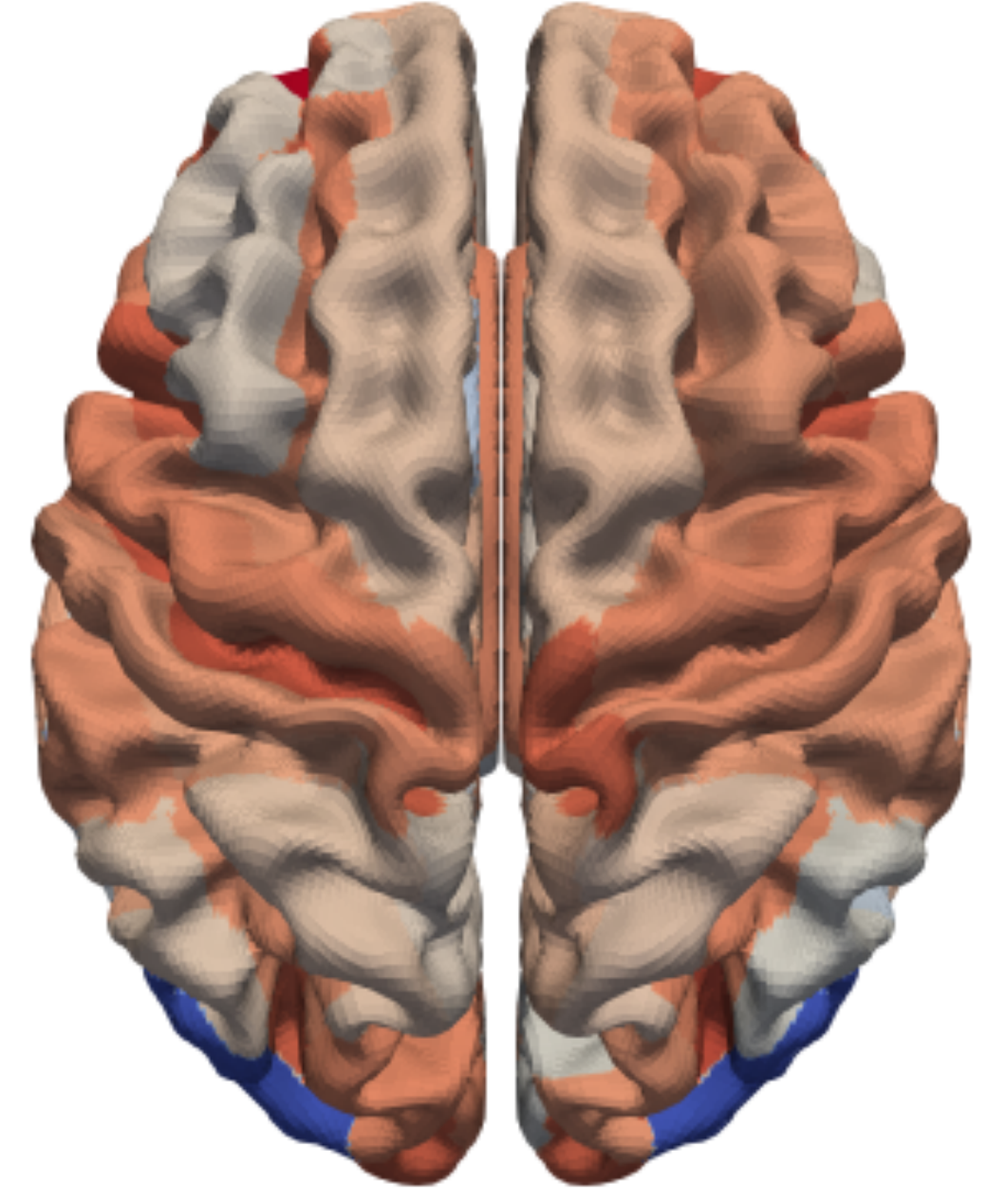}
\caption{Last scan amyloid (left) and thickness (right) as AD}
\label{fig:AE}

\hspace{0.01\textwidth}

\includegraphics[width=0.08\linewidth]{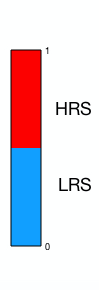}
\includegraphics[width=0.4\linewidth,height=0.44\linewidth]{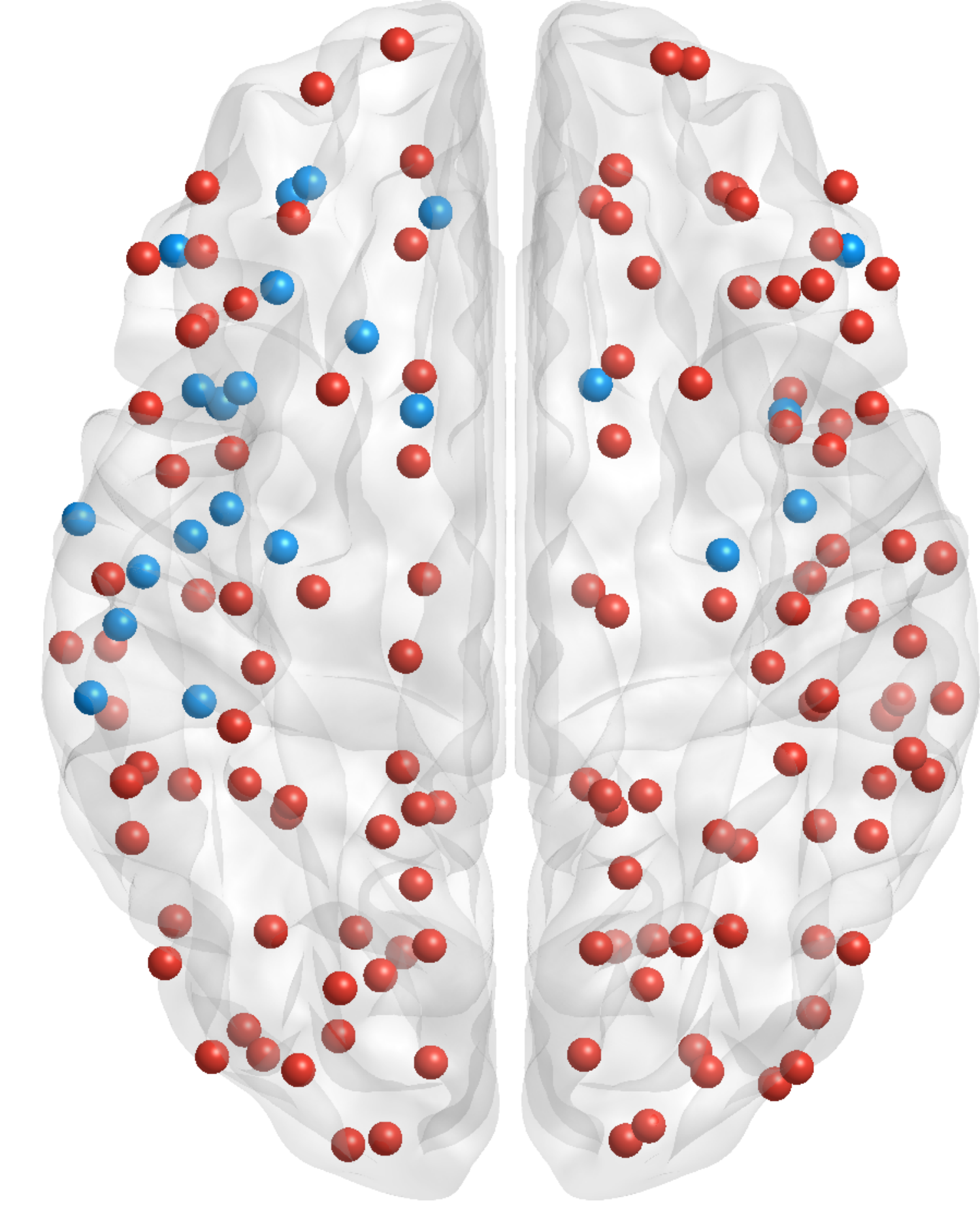}
\hspace{0.01\textwidth}
\includegraphics[width=0.4\linewidth,height=0.44\linewidth]{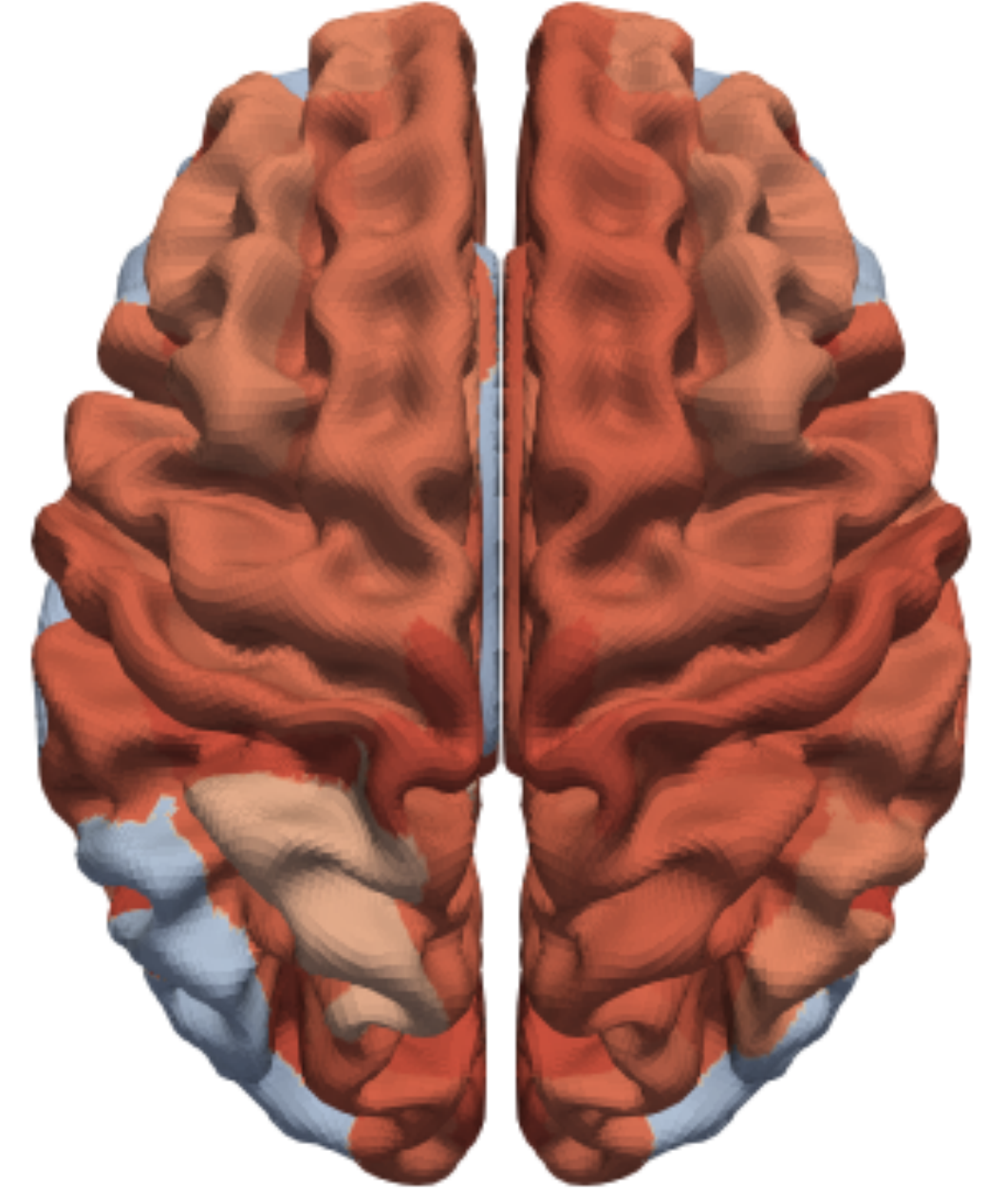}
\hspace{0.05\textwidth}
\caption{Final prediction of amyloid (left) and thickness (right) as AD}
\label{fig:ASIM}
\end{subfigure}
}
\hspace{0.01\textwidth}
\fbox{
\begin{subfigure}{0.46\textwidth}
\centering
\textbf{\large Subject B}

\includegraphics[width=0.4\linewidth,height=0.44\linewidth]{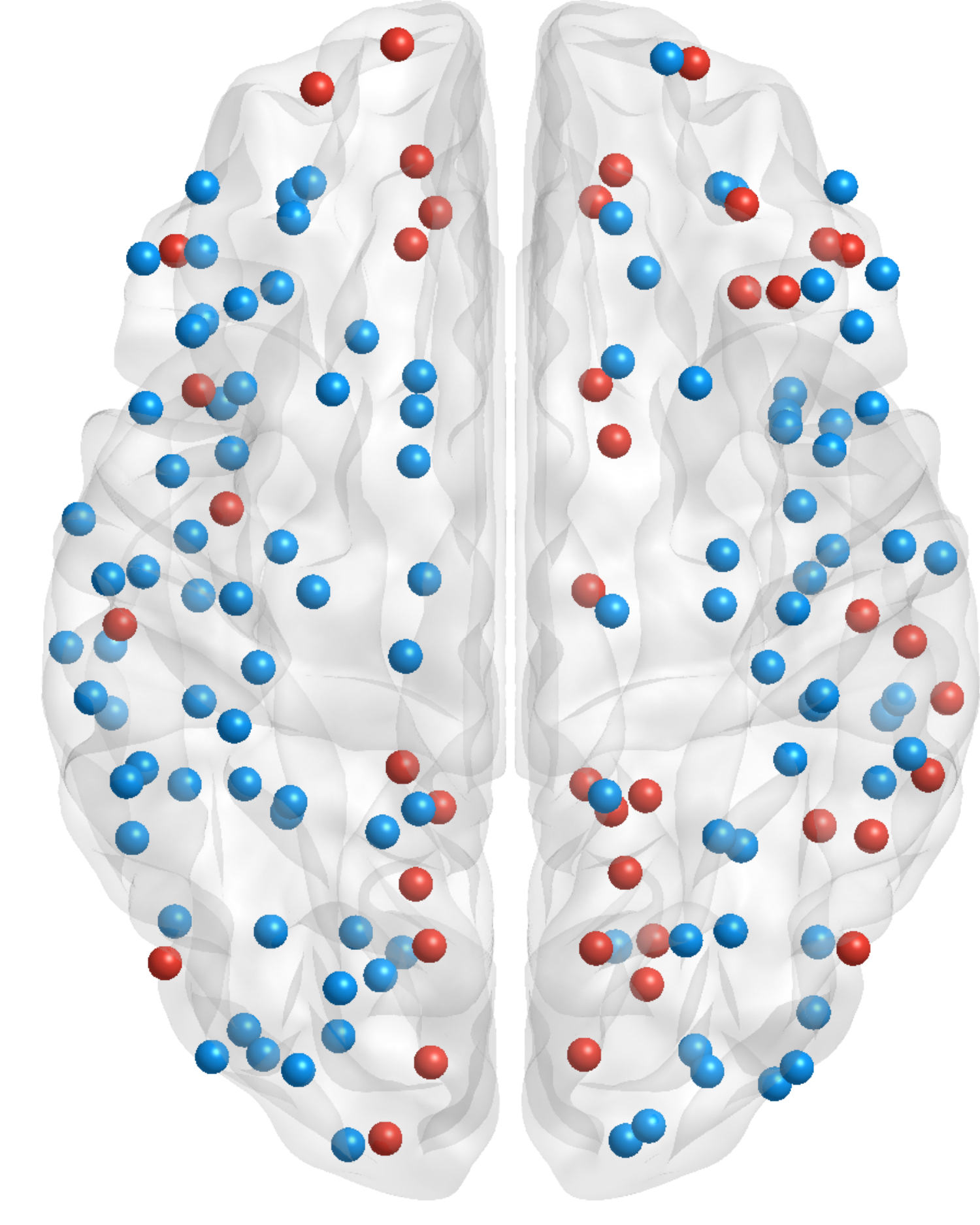}
\hspace{0.01\textwidth}
\includegraphics[width=0.4\linewidth,height=0.44\linewidth]{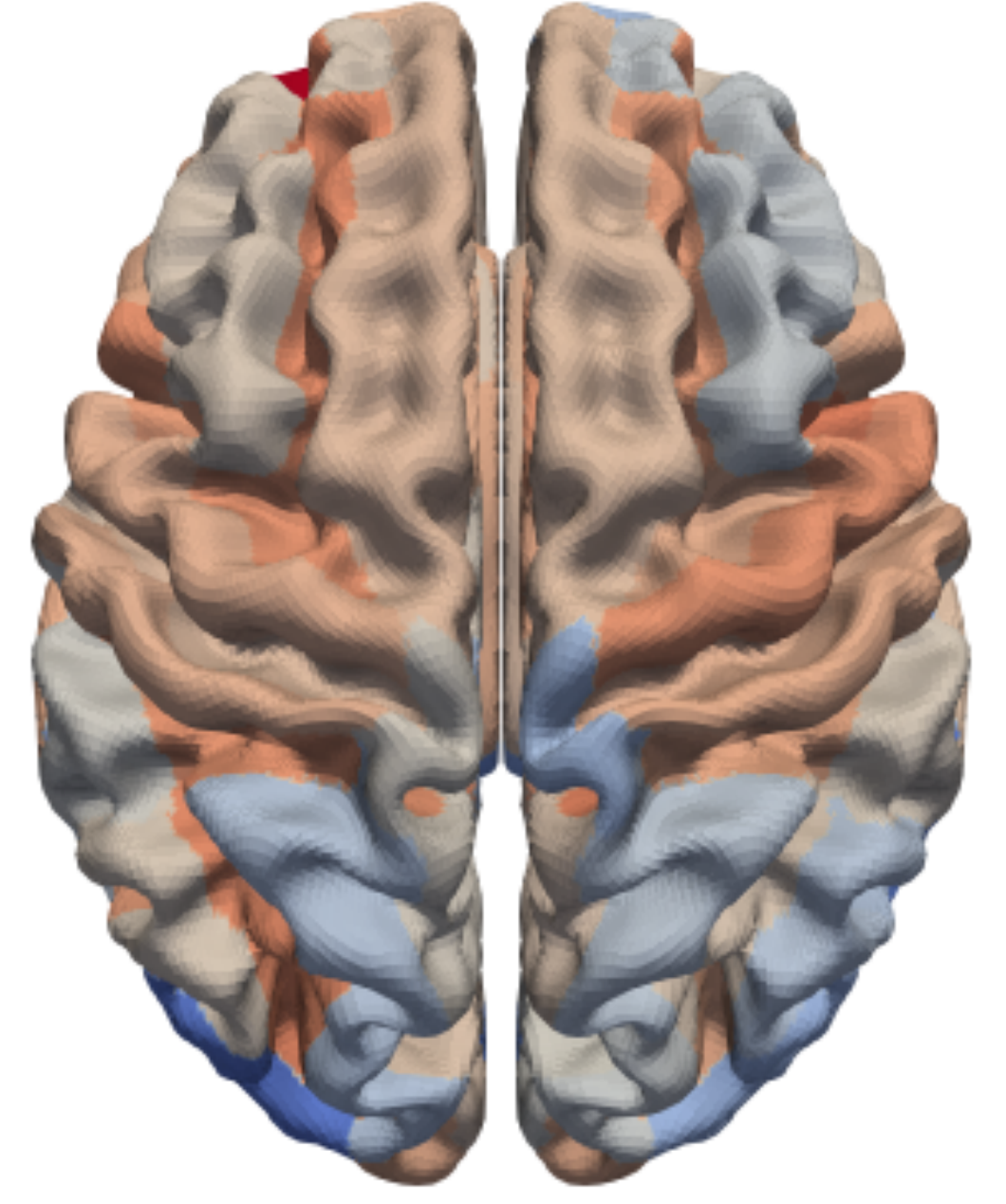}
\caption{Baseline scan of amyloid (left) and thickness (right) as EMCI}

\hspace{0.01\textwidth}

\includegraphics[width=0.4\linewidth,height=0.44\linewidth]{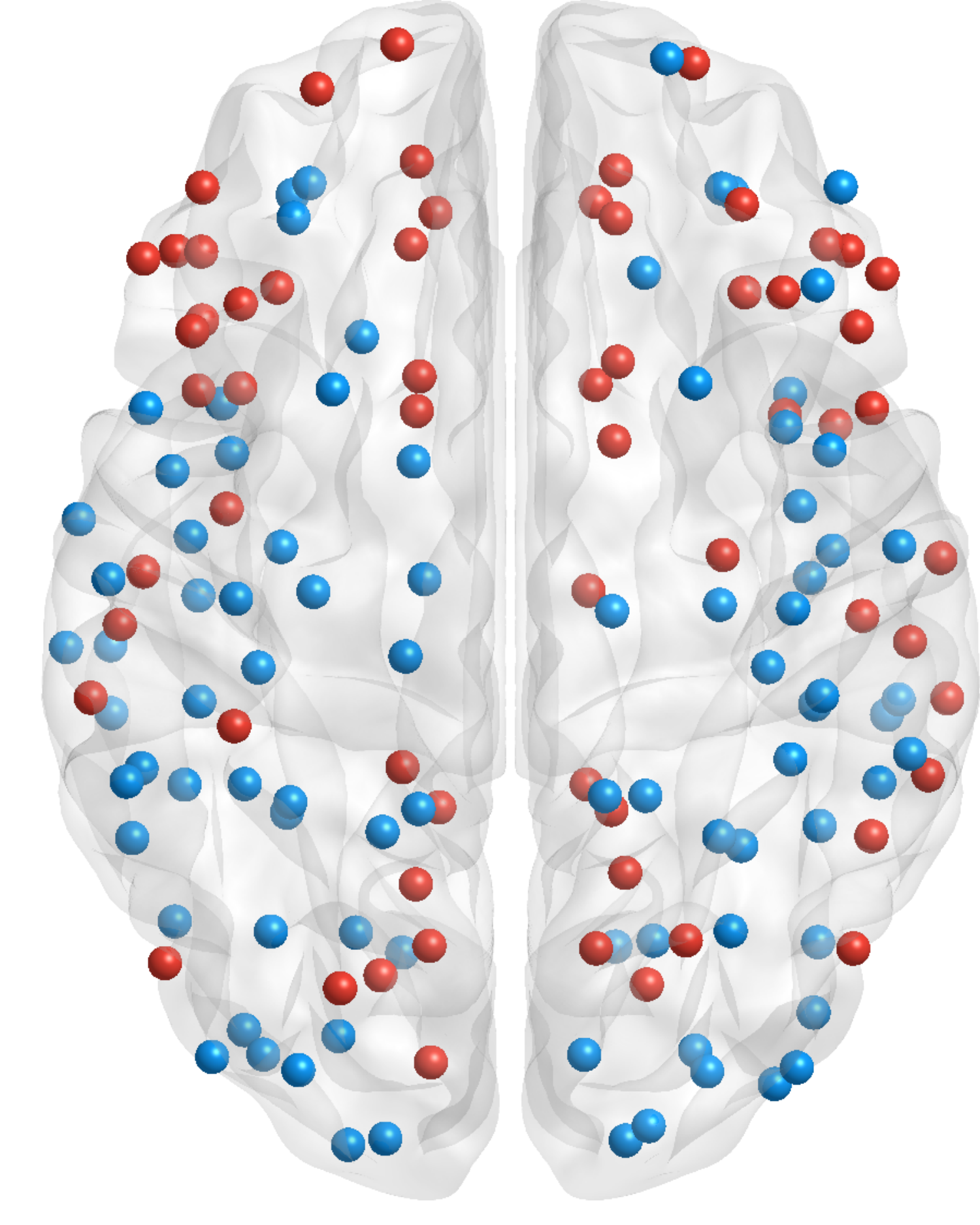}
\hspace{0.01\textwidth}
\includegraphics[width=0.4\linewidth,height=0.44\linewidth]{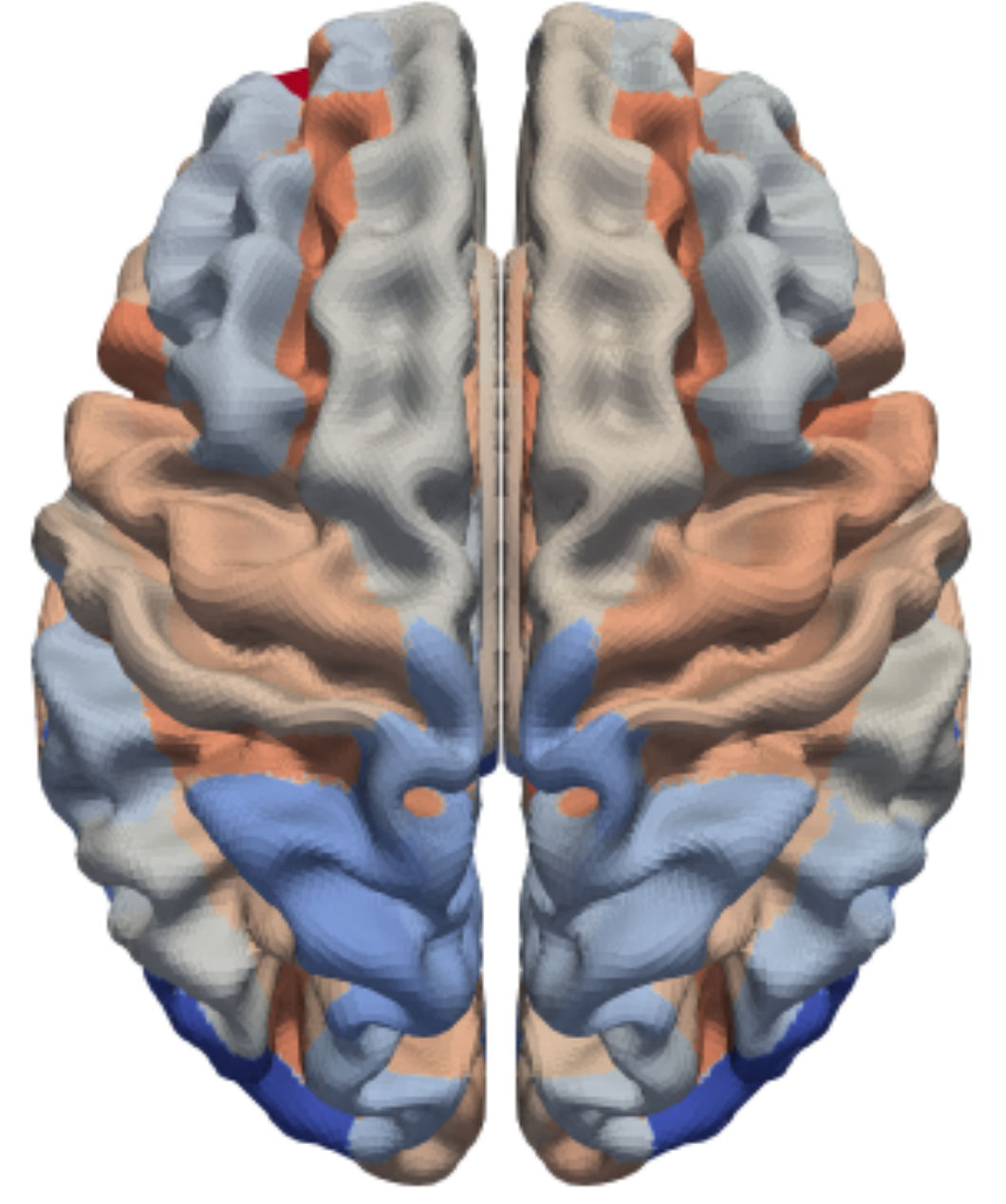}
\caption{Last scan amyloid (left) and thickness (right) as EMCI}

\hspace{0.01\textwidth}

\hspace{0.04\textwidth}
\includegraphics[width=0.4\linewidth,height=0.44\linewidth]{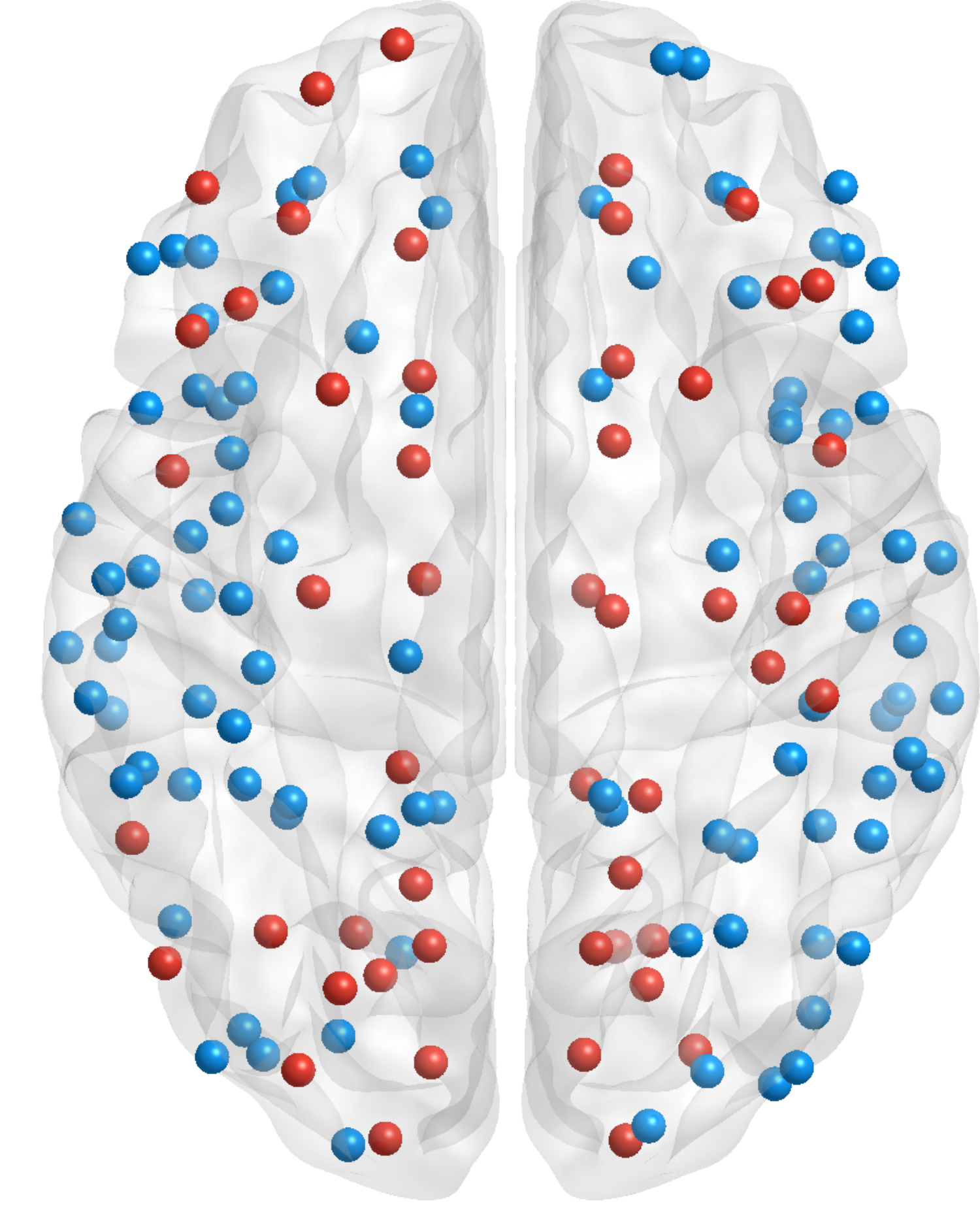}
\hspace{0.01\textwidth}
\includegraphics[width=0.4\linewidth,height=0.44\linewidth]{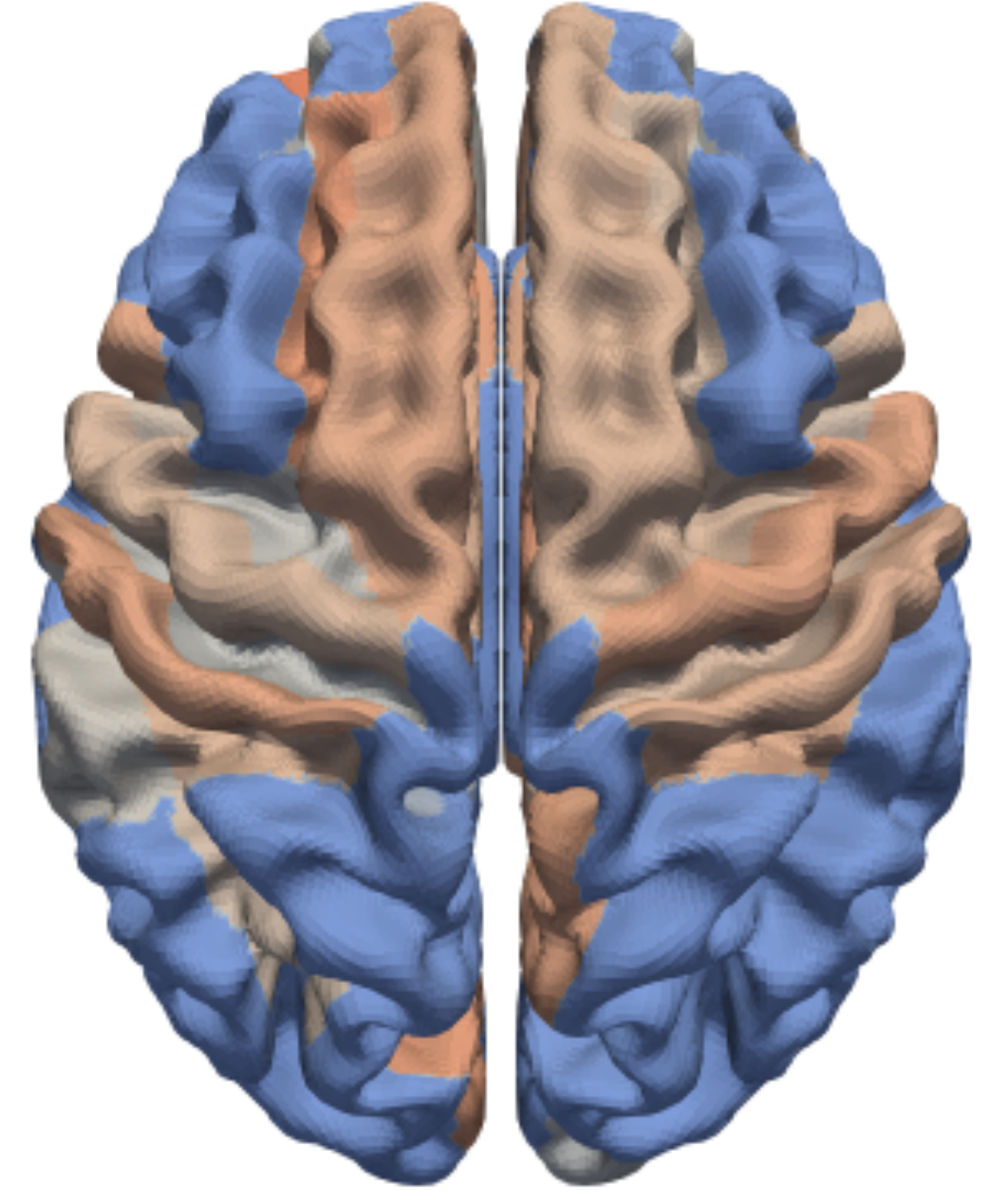}
\includegraphics[width=0.09\linewidth]{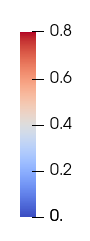}
\caption{Final prediction of amyloid (left) and thickness (right) as CN}
\label{fig:BSIM}
\end{subfigure}
}
\caption{\textbf{Spatiotemporal evolution of amyloid level and cortical thickness of subjects A and B.} 
{ Both subjects were labeled as EMCI in their baseline scans. Subject A was labeled as AD in the last scan and predicted as AD-like, while subject B was labeled as EMCI and predicted as CN-like in the final state by the model.
For the amyloid level distribution, LRS nodes (low amyloid levels) are colored blue and HRS nodes (high amyloid levels) are red; all nodes are one size. MRI scans and corresponding predictions of cortical thickness are surface mapped (via ParaView~\cite{ahrens2005paraview}), where redder colors indicate greater amounts of brain atrophy (low cortical thickness) and bluer areas indicate low- or no- brain atrophy (high cortical thickness). 
}
}
\label{fig:subject}
\end{figure*}

Given a subject's baseline Amyloid PET and MRI imaging data, the model can predict the whole-brain evolution of amyloid deposition and neurodegeneration. Over time, AT[N] will either stay in LRS or develop to HRS, each corresponding to a stable equilibrium. Considering brain atrophy is a complex symptom that could result from a wide range of factors, we set the borderline of AD as 70\%: if more than 104 nodes (70\%) out of 148 nodes reach the HRS, then the subject will be categorized as AD-like, otherwise as CN-like. To compare with simulation results, the average of all subjects' amyloid levels at each node were used to split the empirical amyloid data into low and high states, represented as blue and red nodes shown in Fig.~\ref{fig:subject}.

\textbf{Bistable Classification.}
Bistability enables our model to give a relatively accurate long-term predictive diagnosis, even with similar initial neurological conditions. 
Take subjects A and B as an example (Fig.~\ref{fig:subject}), there is no significant difference in the baseline scans of Amyloid PET and MRI, yet subject A deteriorates into AD while subject B stays in EMCI. In Fig.~\ref{fig:AE}, the last scan of subject A exhibits an obvious increase in A$\beta$ deposition over the brain network. Using the baseline scan as the initial condition, our model successfully predicted the final diagnoses. More than 84.5\% of 148 nodes go to HRS in the final state. A notable increase in brain atrophy was predicted in subject A, showing a consistent pattern between the accumulation of amyloid and degradation of neurons in the simulation results (Fig.~\ref{fig:ASIM}). 
For subject B, while simulation predictions indicated a slight increase in high amyloid nodes across the brain and no significant changes in cortical thickness, the subject is categorized as CN-like in the final state which is consistent with observations. There is no clear sign of neurodegeneration in the last scan of cortical thickness and our model predicts that the majority of the brain regions have a lower risk of neurodegeneration (Fig.~\ref{fig:BSIM}).
As demonstrated above, our model can capture subtle differences in baseline scans between subjects for both A[N] biomarkers. However, there are some brain regions that do not match well with the observed amyloid level. One possible explanation is that in the first and last scans, the average amyloid level is used to separate nodes into low and high states, thus brain regions in the intermediate state (close to the average level) can easily switch between CN-like and AD-like states (blue and red colors, respectively, in Fig.~\ref{fig:subject}). 
Conversely in the simulation, LRS and HRS are distinctly classified by the system's unstable point, referred to as a tipping point.

\begin{figure}[!htb]
\centering
\includegraphics[width=0.4\linewidth]{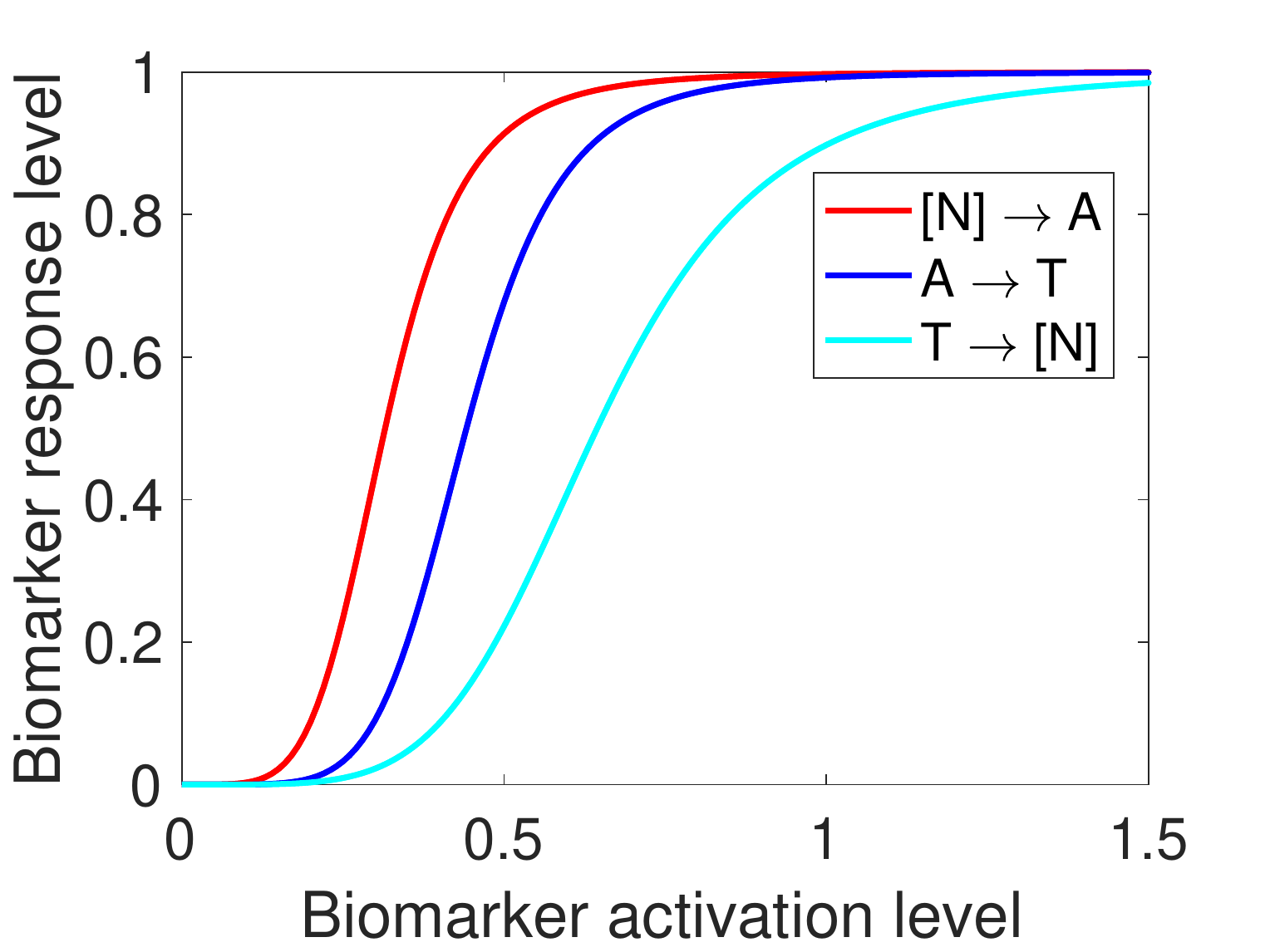}
\caption{\textbf{Interactions between AT[N] biomarkers.} 
{\small The progression of AT[N] biomarkers follow a switch-like pattern, where the three sigmoidal curves represent the activation pathways ($\rm A \rightarrow T$, $\rm A \rightarrow T$) and the positive feedback ($\rm [N] \rightarrow A$), respectively. 
AT[N] slowly accumulates in LRS and increases substantially to HRS once the unstable point is reached. The occurrence of AT[N] follows a sequential cascade proposed in the model.
}
}
\label{fig:interaction}
\end{figure}

\textbf{AT[N] Interactions.} Our model reveals the homologous pattern and sequential occurrence of amyloid deposition, hyperphosphorylated tau, and brain atrophy, which are demonstrated in previous studies~\cite{jack2010hypothetical, sperling2011toward, bertsch2017alzheimer}. In Fig.~\ref{fig:interaction}, we see a cascade of AT[N] development regardless of individual differences, validating the proposed activation-feedback mechanistic pathway. The interactions between AT[N] are individually tested to substantiate the neuropathology. Specifically, 1) A$\beta$, seen as an indicative molecule in the early progression of AD, increases sharply once it reaches the tipping point, i.e., the unstable equilibrium point of the system; 2) following the increase of A$\beta$, tau protein initially stays in the slow accumulative state, then enters a transition period when A$\beta$ passes the tipping point, and finally stabilizes at the high-level state; 3) last, the tau biomarker activates a similar pattern of neurodegeneration. Overall, the switch-like dynamics elicited from the model are in line with AT[N] observations and existing hypothetical models of AD pathology.

\begin{figure*}[!htb]  
\begin{subfigure}{0.22\textwidth}
\centering
\includegraphics[width=0.83\linewidth, height=0.92\linewidth]{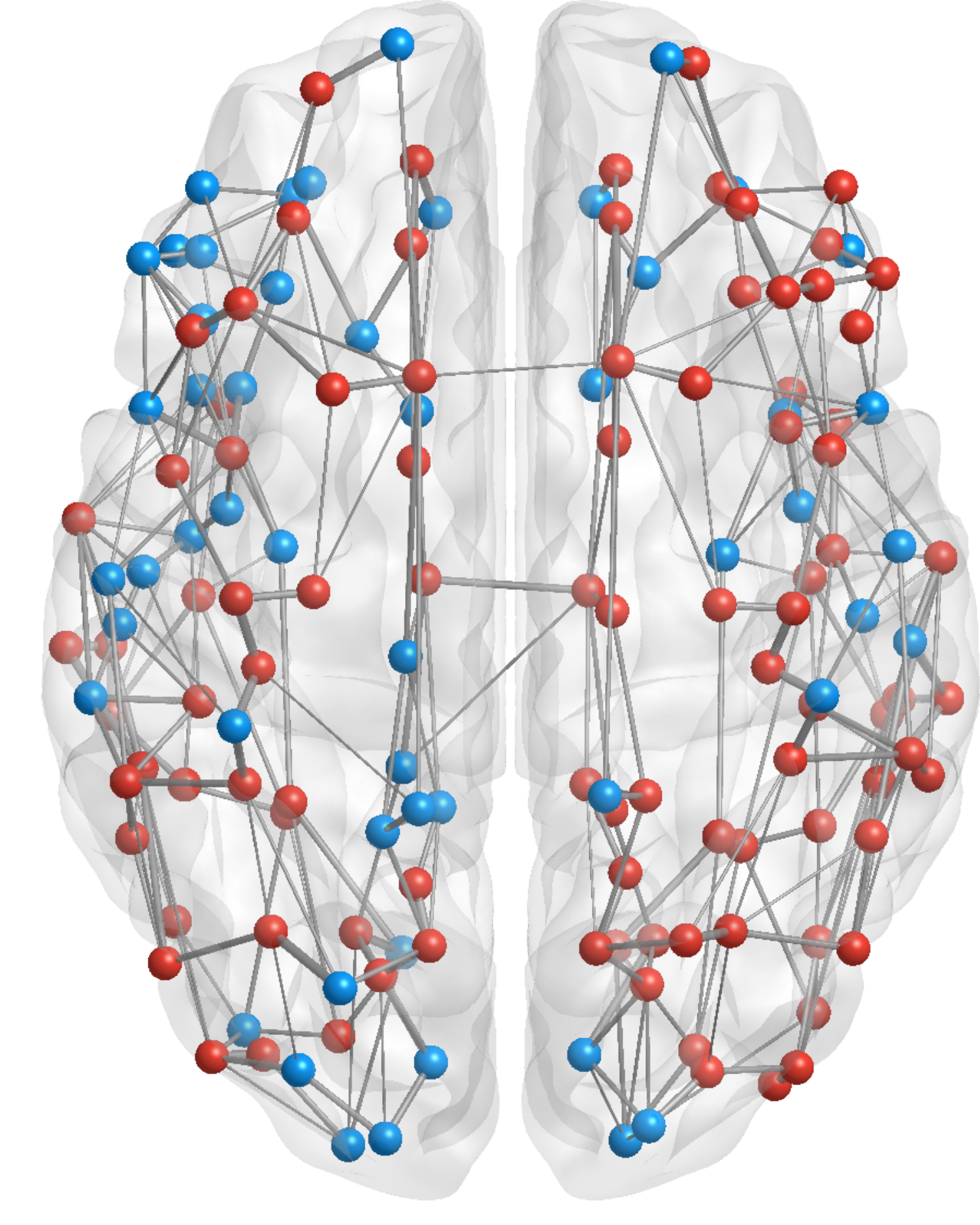}
\caption{$t = 0$}
\label{fig:diff1}
\end{subfigure}
\begin{subfigure}{0.22\textwidth}
\centering
\includegraphics[width=0.83\linewidth, height=0.92\linewidth]{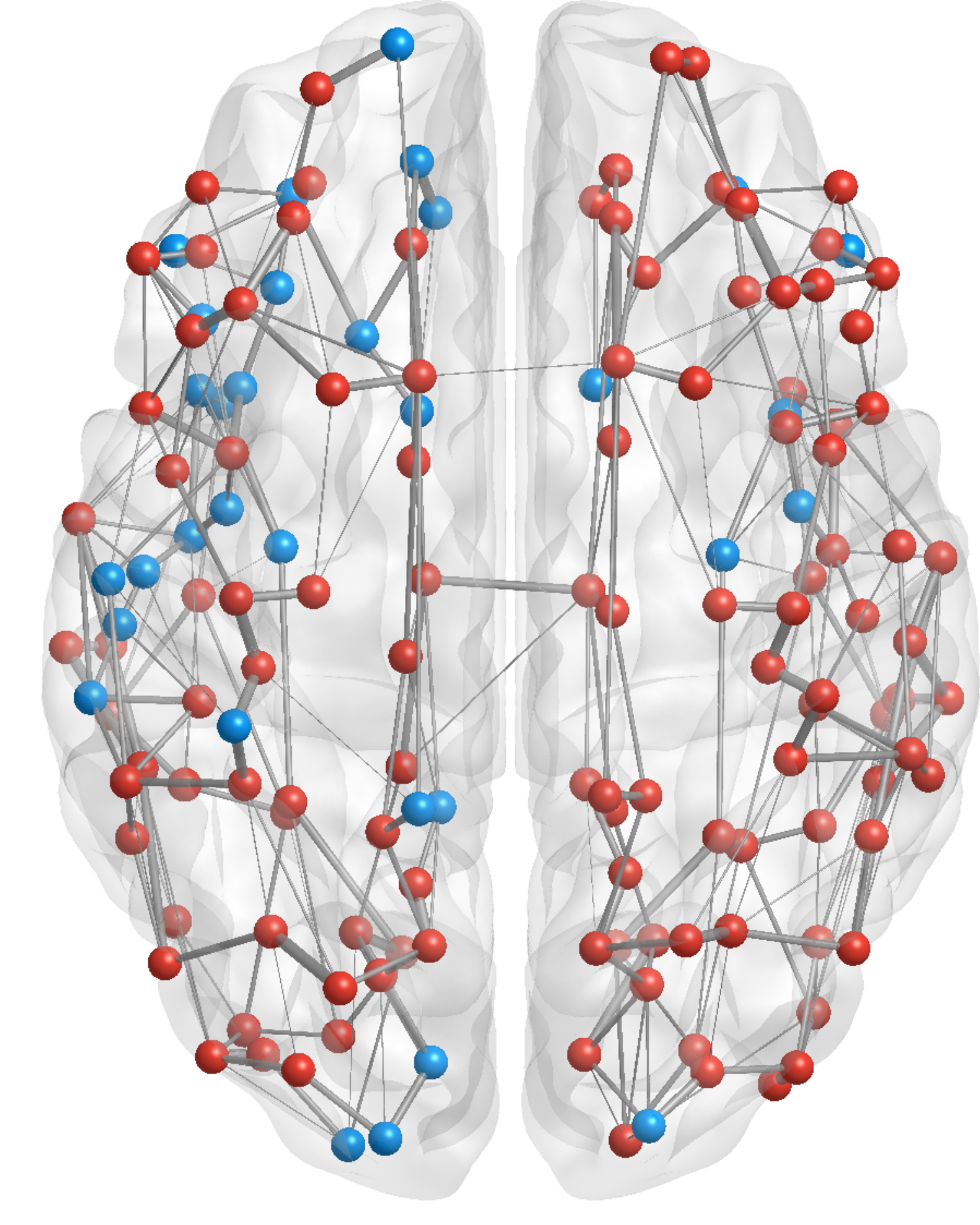}
\caption{$t = 3$}
\label{fig:diff2}
\end{subfigure}
\begin{subfigure}{0.22\textwidth}
\centering
\includegraphics[width=0.83\linewidth, height=0.92\linewidth]{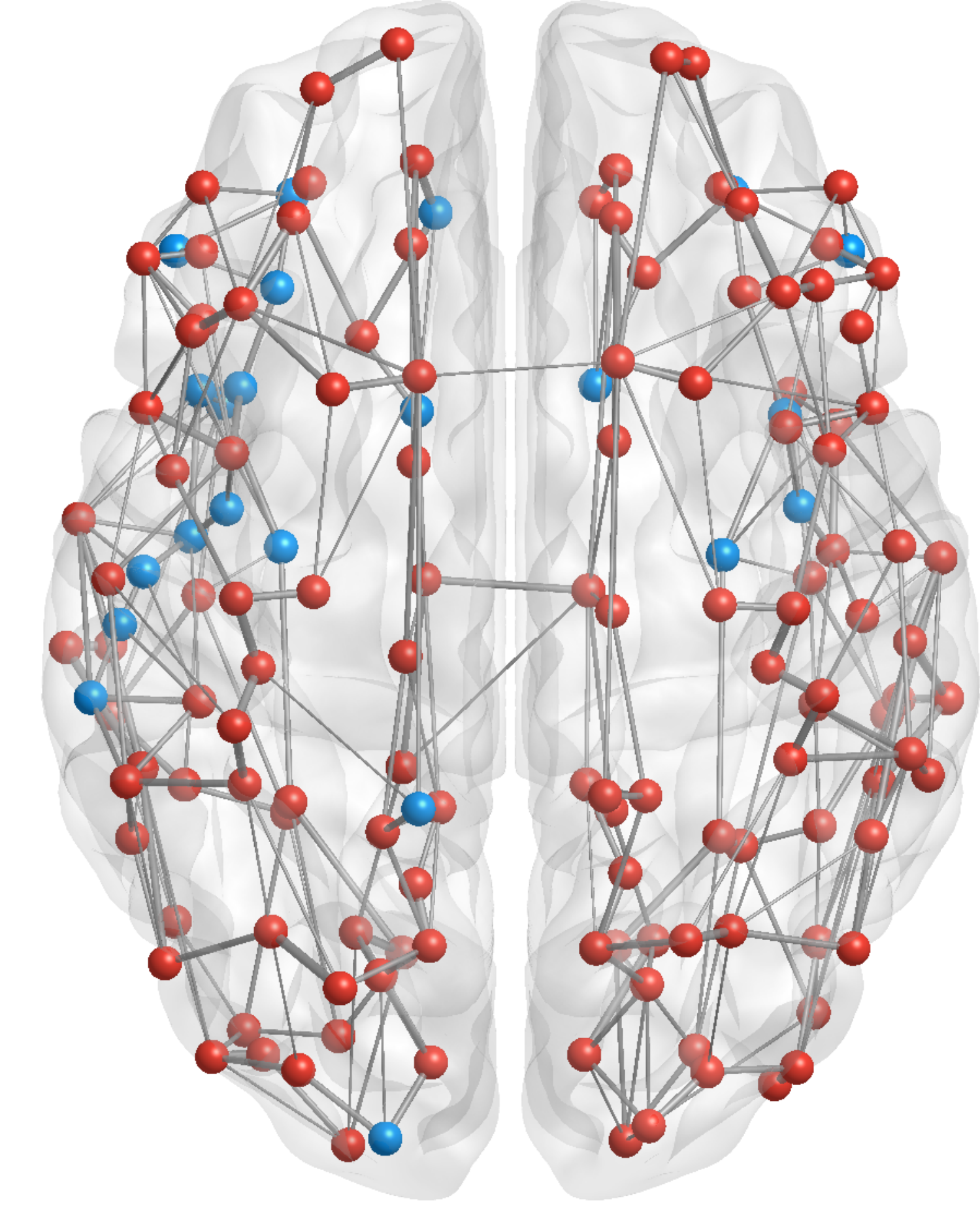}
\caption{$t = 24$}
\label{fig:diff3}
\end{subfigure}
\begin{subfigure}{0.22\textwidth}
\centering
\includegraphics[width=0.83\linewidth, height=0.92\linewidth]{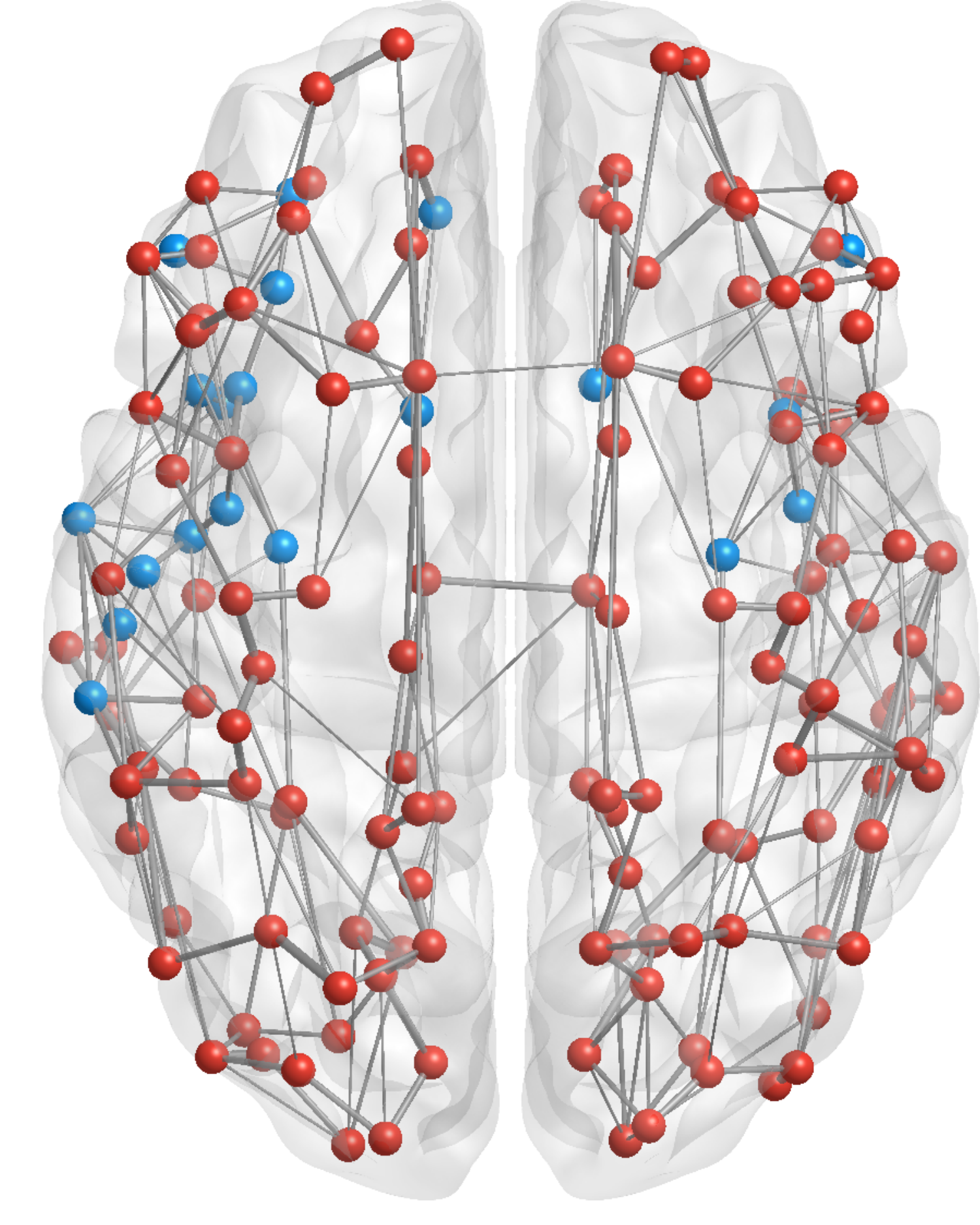}
\caption{$t = 36$}
\label{fig:diff4}
\end{subfigure}
\begin{subfigure}{0.08\textwidth}
\centering
\includegraphics[width=0.6\linewidth]{fig/legendA.jpg}
\end{subfigure}
\caption{\textbf{Amyloid diffusion over the brain network at time $t = 0, 3, 24, 36$ months.} { Beta amyloid contagion-like behavior and spread to adjacent nodes through the brain network.}
}
\label{fig:diff}
\end{figure*}

\textbf{Network Diffusion.} In this model, we use the average network and connectivity strength of 506 subjects calculated from DTI scans. Network connectivity strength is used in the diffusion term to characterize the diffusion strength. 
Closer examination of amyloid diffusion in the system (Fig.~\ref{fig:diff}) presents the spatial and temporal dynamics of amyloid levels across the brain. For example, when $t=0$, inferior occipital gyrus (node 2) stays in LRS and middle occipital gyrus (the neighboring node 19) is in HRS. However, as time proceeds, A$\beta$ diffuses on the network from node 19 to node 2, increasing the amyloid level at node 2 and finally transitioning it to HRS. After 36 months, 32 nodes in total change from LRS to HRS, demonstrating the prion-like progression of amyloid. 


\textbf{Parameter Optimization.}
To find the best parameter vector for our model, we define the objective function, $f(\Theta)$, as the number of wrong categorizations over the 200 cohorts. Thus the optimization problem is
\begin{equation*}\label{eq:loss}
    \min_{L\le \Theta \le U}f(\Theta),
\end{equation*}
where $[L, U]$ is a searching box in $\mathbb{R}^{16}$ defining the feasible set. A rough estimation of the lower and upper boundaries for model parameters are provided in Table~\ref{tab:para}, where most parameter values sampled in the space captured the bistable feature. Using the genetic algorithm to optimize model parameters in the searching box, the best set of parameter values found are shown in Table~\ref{tab:para}. 
Our model can achieve a classification accuracy of 83.0\% if we assume the diagnosed label from last scan is the subject's final projection. Varying individual parameters, we further analyze the sensitivity of model parameters by measuring the changes in system's final states and classification outcomes. It is found that our model is sensitive to the degradation and diffusion of amyloid and tau protein, network resilience, and the positive feedback; conversely, the three hill coefficients and T[N] activation rate constant play a weak role in model dynamics. 


\begin{table*}[!htb]
\centering
\caption{Best found values, ranges, sensitivities of model parameters.}
\begin{tabular}{ l l l l | l l l l } 
\toprule
\toprule
 Parameter & Best Value & Range & Sensitivity & Parameter & Best Value & Range & Sensitivity \\ 
 \midrule
$k_{pA}$ & 0.20 & [0.06 1.00] & Moderate & $k_{MN}$ & 0.32 & [0.12 0.37] & Sensitive \\
$k_{pT}$ & 0.53 & [0.22 1.78] & Moderate & $k_{MA}$ & 0.44 & [0.16 0.45] & Moderate  \\ 
$k_{dA}$ & 2.87 & [0.54 3.09] & Sensitive & $k_{MT}$ & 2.49 & [0.20 0.67] & weak \\ 
$k_{dT}$ & 4.07 & [1.36 4.64] & Sensitive & $\alpha$ & 5.11 & [2.92 10.00] & weak  \\ 
$k_{NA}$ & 1.47 & [1.36 8.00] & Moderate & $\beta$  & 5.97  & [3.16 8.58] & weak \\ 
$k_{AT}$ & 2.59 & [2.51 8.00] & Moderate & $\gamma$ & 4.94 & [1.16 5.84] & weak \\ 
$k_{TN}$ & 3.85 & [2.92 8.00] & Moderate & $d_A$  & 0.15 & [0.01 1.00] & Sensitive \\ 
$k_{r}$  & 2.49 & [0.1 4.26] & Sensitive & $d_T$  & 0.78 & [0.01 3.00] & Sensitive \\ 
\bottomrule
\bottomrule
\end{tabular}
\label{tab:para}
\end{table*}

\begin{figure}[!htb]
\centering
\begin{subfigure}{0.4\textwidth}
\centering
\includegraphics[width=1\linewidth]{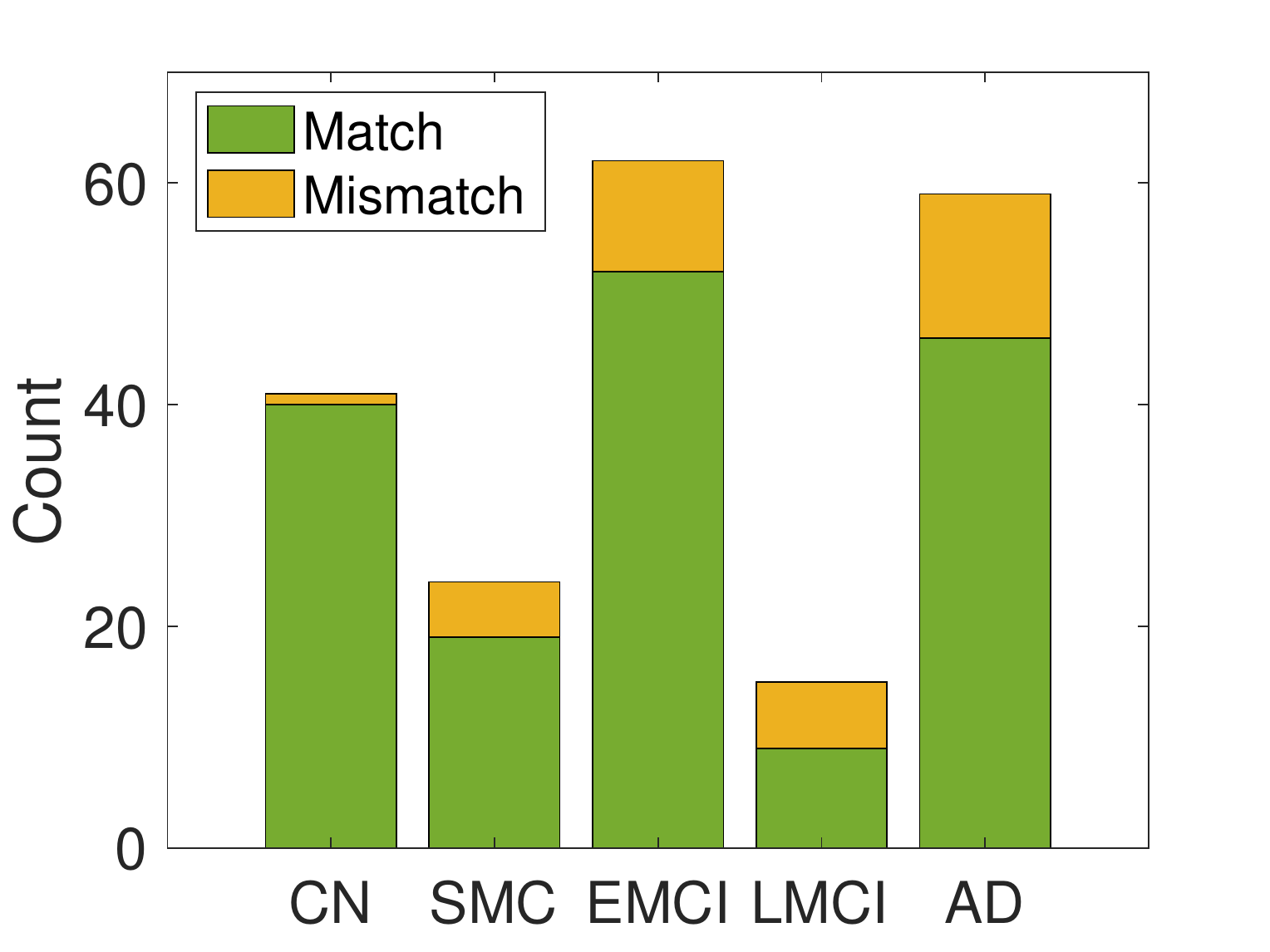}
\caption{Match vs. Mismatch}
\label{fig:mismatch_label}
\end{subfigure}
\begin{subfigure}{0.4\textwidth}
\centering
\includegraphics[width=1\linewidth]{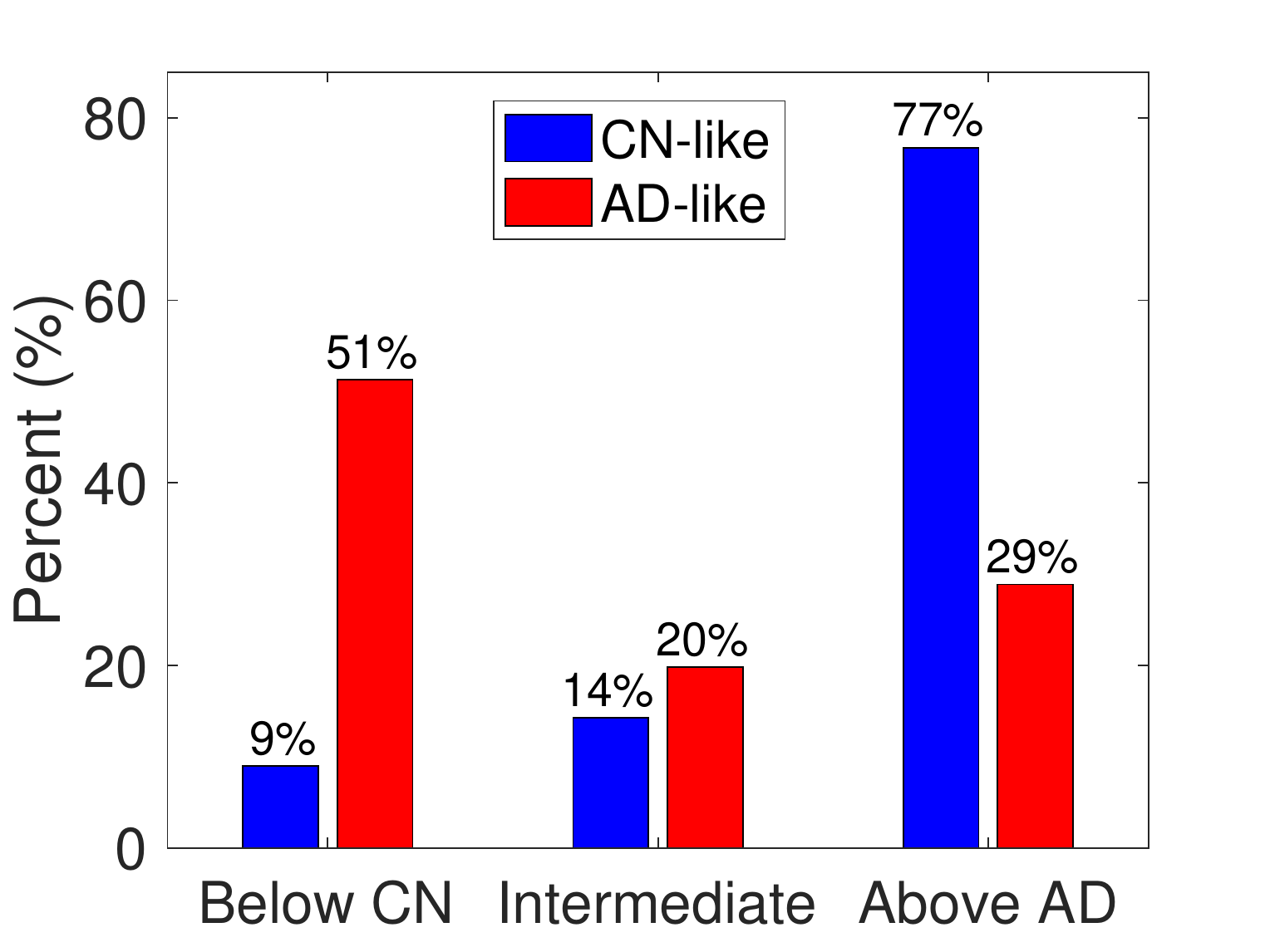}
\caption{Amyloid categorization}
\label{fig:mismatch_percent}
\end{subfigure}
\caption{\textbf{Misclassification analysis.} { \textbf{(a)} Mismatch count in five diagnostic labels with green and yellow representing correct and wrong predictions, respectively. \textbf{(b)} Average percent of nodes from misclassified patients partitioned into three states: below the average amyloid level of CN-like group at each node, intermediate, and above the average amyloid level of AD-like group at each node. Blue bars show CN-like subjects misclassified as AD and red bars show AD-like subjects misclassified as CN.}
}
\label{fig:mismatch}
\end{figure}

\textbf{Misclassification Analysis.}
Among the 17\% misclassified subjects, 12.6\% of CN-like subjects were mistakenly classified as AD and 26.0\% of AD-like subjects were misclassified as CN. In practice, the overall prediction error could potentially be lower than 17\% considering that some CN-like subjects may later develop AD.
Fig.~\ref{fig:mismatch_label} shows the misclassification over five diagnostic groups, where around 43\% percent of the total mismatches are from the EMCI and LMCI groups. This could be partially explained by the blurred demarcation between the symptoms of EMCI and LMCI in clinical diagnosis. 
Given the massive overlap of both amyloid level and cortical thickness level between CN-like and AD-like groups (as seen in Figs.~\ref{fig:amyloidHist} and~\ref{fig:corHist}), it is challenging to achieve high prediction accuracy. To address this, we calculated the mean amyloid levels separately for CN-like and AD-like subjects by node, resulting in a mean amyloid level for each group at each node to be used as a threshold for categorizing amyloid levels from all misclassified subjects. Using misclassified subjects' last scans, we calculated the average percentage of nodes with amyloid levels falling into three categories: below the average of CN-like group, intermediate, above the average of AD-like group. This paradox is exemplified in  Fig.~\ref{fig:mismatch_percent}, where subjects were labeled as CN even though 77\% of brain regions have higher amyloid levels than the AD-like group mean, when by intuition it is more likely to be recognized as AD (which was the prediction of our simulation). A similar pattern exists in the misclassified AD group where 51\% of nodes have lower amyloid levels than the CN-like group mean. These anomalous empirical data from misclassified subjects may result from individual differences, precision limits in scanning technology, errors in neuroimaging processing, or variations in node-specific biomarkers' evolution patterns. With the inclusion of tau PET data in the future, we could further improve the prediction power of our model and optimize model parameter.





\section{Conclusion}
In this work, we propose a novel network-guided systems biology approach to investigate the reaction-diffusion process of AT[N] biomarkers throughout the brain, which provides us with a system-level underpinning of the physiopathological mechanism in AD. 
Our integrated method across systems biology and neuroscience is a pioneer computational solution that jointly models the dynamics of observed AT[N] biomarkers on the setting of the brain network. Second, we also include brain resilience as a moderator in our model. Thus, our model can quantify the level of network resilience to AD pathology in individual regions across the brain. Model results demonstrate the ability to capture presaging information from neuroimaging data and the prediction of long-term subject-specific AD progression pattern. 

In the next step, we plan to incorporate Tau PET scans from the 200 cohorts into our model. Provided with all three biomarkers' neuroimaging data, we can further validate model results and increase model versatility. An advanced high-dimensional optimization algorithm will be used to optimize system parameters to improve the overall prediction power. 





\bibliographystyle{unsrt}  
\bibliography{template}
\end{document}